\documentclass[12pt]{revtex4}
\usepackage{amssymb}
\usepackage{epsfig,newlfont}
\usepackage{subfigure}
\usepackage[colorlinks, 
linkcolor=blue,
urlcolor=blue,
citecolor=blue,
bookmarksopen=false]{hyperref}
\usepackage{graphicx, xcolor}
\usepackage{caption}

\def\newblock{\hskip .11em plus .33em minus .07em}
\newcommand{\changes}[1]{{#1}}

\begin{document}

\title{Moderation of neoclassical impurity accumulation in high temperature plasmas of helical devices}
\author{J.L. Velasco, I. Calvo, S. Satake$^{2,3}$, A. Alonso, M. Nunami$^{2,3}$, M. Yokoyama$^{2,3}$, M. Sato$^2$, T. Estrada, J. M. Fontdecaba, M. Liniers, K. J. McCarthy, F. Medina, B. Ph Van Milligen, M. Ochando, F. Parra$^{4,5}$, H. Sugama$^2$, A. Zhezhera$^6$, {the LHD experimental team}, {the TJ-II team}}

\affiliation{Laboratorio Nacional de Fusi\'on, CIEMAT, Madrid, Spain}
\affiliation{$^2$National Institute for Fusion Science, Toki, Japan}
\affiliation{$^3$The Graduate School for Advanced Studies (SOKENDAI), Toki, Japan}
\affiliation{$^4$Rudolf Peierls Centre for Theoretical Physics, University of Oxford, Oxford, UK}
\affiliation{$^5$Culham Centre for Fusion Energy, Abingdon, UK}
\affiliation{$^6$Institute of Plasma Physics, NSC KIPT, Kharkov, Ukraine}

\begin{abstract}

Achieving impurity and helium ash control is a crucial issue in the path towards fusion-grade magnetic confinement devices, and this is particularly the case of helical reactors, whose low-collisionality ion-root operation scenarios usually display a negative radial electric field which is expected to cause inwards impurity pinch. In this work we discuss, based on experimental measurements and standard predictions of neoclassical theory, how plasmas of very low ion collisionality, similar to those observed in the \textit{impurity hole} of the Large Helical Device~\cite{yoshinuma2009observation,ida2009observation,yokoyama2002lhd}, can be an exception to this general rule, and how a negative radial electric field can coexist with an outward impurity flux. This interpretation is supported by comparison with documented discharges available in the International Stellarator-Heliotron Profile Database, and it can be extrapolated to show that achievement of high ion temperature in the core of helical devices is not fundamentally incompatible with low core impurity content.

\end{abstract}

\maketitle

\section{Introduction}

In magnetic confinement fusion, atomic species other than the fusion reactants (e.g. deuterium and tritium for the fuel mix envisaged for the first demonstration reactors) are termed "€œimpurities"€. The presence of even small concentration of impurities (especially those of high charge number $Z$) in the confinement volume has deleterious consequences on plasma performance, due to radiation power loss and fuel dilution. Furthermore, their accumulation in the core region can ultimately preclude the steady state operation of a fusion reactor. On the grounds of the standard neoclassical theory, such accumulation is expected to occur for medium and high-$Z$ impurities. \changes{This is particularly the case of stellarator-type reactors, for which a negative radial electric field (i.e. directed towards the core) is a natural condition, and no ion temperature screening effect is expected (see e.g.~\cite{dinklage2013ncval,regana2013euterpe,igitkhanov2006impurity}), unlike in the tokamak configuration}.  Finally, the use of heavy species such as tungsten as the divertor and first wall material is, to date, the preferred option to meet the requirements of heat exhaust, material erosion, high radiation fraction and tritium retention.

In view of these facts, a robust strategy for the control of core impurity accumulation is needed as part of an integral solution to the several requirements to achieve magnetic confinement fusion. The two fundamental approaches are (1) to control the source of impurities from the divertor and plasma facing components by tailoring of the scrape-off layer (SOL) regime and (2) to act on the radial transport of impurities in the confined region. However, it should be noted that stringent conditions on SOL and core regimes are imposed by detachment and fusion performance respectively.

Plasma discharge scenarios with controlled core impurity concentration have been demonstrated in several devices. Central heating with microwaves in the electron and ion cyclotron frequencies has been shown to be instrumental for controlling impurities in tokamaks~\cite{neu2002aug,doyle2007iter}. In helical devices, experimental conditions with low impurity confinement time and very low impurity concentration levels in the core have been documented: the High Density H (HDH) mode in the stellarator Wendelstein 7-AS~\cite{mccormick2002hdh} and the \textit{impurity hole} in high-temperature low-density plasmas of the Large Helical Device (LHD)~\cite{yoshinuma2009observation,ida2009observation}.

In LHD, an internal ion energy transport barrier can be developed in low-density ($\sim 10^{19}\,$m$^{-3}$) plasmas heated with Neutral Beam Injection (see e.g.~\cite{yokoyama2008itb}). This leads to high ion temperature (around 5$\,$keV) with a steep gradient, together with electron temperatures that are usually slightly lower (around 3 --- 4 $\,$keV). After the transport barrier has formed, the impurity density decreases~\cite{ida2009observation,yoshinuma2009observation}, faster in the core than at outer positions, hence a positive impurity density gradient develops in the region of high ion temperature inside the transport barrier. This is the so-called impurity hole, which is most easily achieved in the neoclassically unoptimized outwards-shifted magnetic configuration of LHD~\cite{beidler2011ICNTS}. The impurity content reduction is seen to be larger for neon ($Z\!=\!10$) than for carbon ($Z\!=\!6$) and helium ($Z\!=\!2$)~\cite{yoshinuma2010species}, and Soft-x-ray emission indicates that high-Z species are also affected~\cite{ida2009observation}. While in plasmas of smaller temperature, positive radial electric fields have been measured in the edge region and connected to changes in impurity transport~\cite{nakamura2014shielding}, impurity hole plasmas display a negative electric field in the core~\cite{ido2010hibp}; in spite of this experimental observation, transport analyses need an outward pinch in order to describe the time evolution of the impurity concentration. 

Finally, core electron root confinement (CERC)~\cite{yokoyama2007cerc}, a regime observed in several helical devices and characterized by a large and positive radial electric field, is also expected to show reduced impurity content; nevertheless, it is usually not considered reactor-relevant, as it is most easily obtained when the electron temperature is much higher than the ion temperature, a situation that cannot be achieved when the thermal coupling between species is strong. In a previous work~\cite{yokoyama2002lhd}, neoclassical transport analyses were performed with artificial LHD plasmas of high ion temperature, in order to study the possibility to access electron root and improved energy confinement. An experimental confirmation of this prediction is LHD plasmas with electron and ion internal transport barrier~\cite{nagaoka2015etb}: those plasmas are characterized by very high electron and ion core temperature; the measured radial electric field is positive, and a positive electron root is predicted by neoclassical theory.

In this work we study the neoclassical properties of plasmas with very low ion collisionality that have not made a transition to electron root, focusing in the potential impact on impurity transport. The radial electric field of these plasmas, although negative, causes a thermodynamical force not much larger than the ion temperature gradient, which opens the possibility that additional outward pinches (associated e.g. to asymmetries in the impurity distribution on the flux-surface~\cite{regana2013euterpe,alonso2016inertia} or to turbulent mechanisms~\cite{mikkelsen2014gk,nunami2016iaea}) become relevant and overcome the small inward neoclassical pinch. We will show that LHD impurity hole plasmas are an example of this scenario, which we will characterize as well with new experiments in low density plasmas of the heliac TJ-II~\cite{sanchez2013tj-ii} created and heated by means of Neutral Beam Injection. We will further support our conclusions comparing these plasmas with documented discharges from the International Stellarator-Heliotron Profile Database (ISHPDB)~\footnote{\url{https://ishpdb.ipp-hgw.mpg.de/}\\\url{http://ishpdb.nifs.ac.jp/index.html}} and we will discuss the extrapolation to larger devices.

The paper is organized as follows: the equations that determine the radial electric field in helical devices are presented in section~\ref{SEC_EQ}. The solution of small radial electric field and its parameter dependence are discussed in section~\ref{SEC_SCA}. The general predictions are compared to simulations for LHD in section~\ref{SEC_LHD} and to experiments at TJ-II in section~\ref{SEC_TJ-II}. We summarize the results in section~\ref{SEC_DIS}.

\section{Equations}\label{SEC_EQ}

According to standard neoclassical theory, the particle flux of species $b$ per unit area through a magnetic surface, $\Gamma_b$, can be written for low collisionalities as:
\begin{equation}
 \frac{\Gamma_b}{n_b}=- L_{11}^b\left(\frac{n_b'}{n_b}- \frac{Z_beE_r}{T_b}+\delta_b\frac{T_b'}{T_b}\right)\,,
\label{EQ_LINEAR}
\end{equation}
where $Z_be$ is the charge of the species, $n_b$ and $T_b$ are the density and temperature, $L_{11}^b$ and $\delta_b$ are the neoclassical thermal transport coefficients (with their usual definition, see e.g.~\cite{beidler2011ICNTS,maassberg1999densitycontrol}), and prime stands for radial derivative. We will use $r$ as radial coordinate: $V\!=\!2\pi^2 Rr^2$ will be the volume enclosed by the flux-surface denoted by $r$, with $R$ being the major radius. The radial electric field $E_r$ is set by ambipolarity of the neoclassical fluxes {(generally speaking, this provides an accurate quantitative prediction of $E_r$ in the core region of helical devices, while closer to the edge the agreement with the experiment is typically qualitative~\cite{dinklage2013ncval})}. If, for the sake of simplicity, we consider a plasma with trace impurities ($Z_in_i=n_e\gg Z_In_I$), the ambipolar equation reads:
\begin{eqnarray}
- L_{11}^e\left(\frac{n_e'}{n_e}+ \frac{eE_r}{T_e}+\delta_e\frac{T_e'}{T_e}\right) = - L_{11}^i\left(\frac{n_e'}{n_e}- \frac{Z_ieE_r}{T_i}+\delta_i\frac{T_i'}{T_i}\right)\,.\label{EQ_AMB}
\end{eqnarray}
Although the transport coefficients generally depend on $E_r$, as we will explicitly discuss in the next section, it is illustrative to rewrite equation~(\ref{EQ_AMB}) as:
\begin{equation}
\frac{eE_r}{T_i} = \frac{L_{11}^i-L_{11}^e}{Z_iL_{11}^i+\frac{T_i}{T_e}L_{11}^e}\frac{n_e'}{n_e} + \frac{\delta_iL_{11}^i\frac{T_i'}{T_i} - \delta_eL_{11}^e\frac{T_e'}{T_e}}{Z_iL_{11}^i+\frac{T_i}{T_e}L_{11}^e}\,.\label{EQ_ER}
\end{equation}
If we now introduce this expression into the radial impurity flux, we have:
\begin{equation}
 \frac{\Gamma_I}{n_I}=- L_{11}^I\frac{n_I'}{n_I} + Z_IL_{11}^I \frac{\delta_iL_{11}^i\frac{T_i'}{T_i} - \delta_eL_{11}^e\frac{T_e'}{T_e}}{Z_iL_{11}^i+\frac{T_i}{T_e}L_{11}^e} -  \delta_I L_{11}^I\frac{T_i'}{T_i}\,,\label{EQ_FLUX}
\end{equation}
where for simplicity we have neglected terms proportional to the electron density gradient (a reasonable approximation when the electron density profile is rather flat) and assumed that the impurities thermalize with the bulk ions, hence $T_I=T_i$. In ion-root plasmas, it is usually reasonable to assume (see e.g.~\cite{igitkhanov2006impurity}) that, due to the much smaller electron Larmor radius, for similar electron and ion temperatures we have $L_{11}^e\ll L_{11}^i$, and equation~(\ref{EQ_FLUX}) can be simplified to:
\begin{equation}
 \frac{\Gamma_I}{n_I}=- L_{11}^I\frac{n_I'}{n_I} + L_{11}^I \left(\frac{Z_I}{Z_i}\delta_i - \delta_I\right)\frac{T_i'}{T_i}\,.\label{EQ_SIMP}
\end{equation}

It can be easily shown~\cite{maassberg1999densitycontrol} that $\delta_b$ is approximately equal to 1/2, 5/4, 7/2 and 3/2 if species $b$ is in pure $\nu$, $\sqrt{\nu}$, $1/\nu$, and plateau~\cite{igitkhanov2006impurity} neoclassical collisionality regimes respectively, independently of the magnetic configuration and species (note however that impurities of higher $Z_I$ will be in more collisional regimes, and that the collisionality regime, for a given density and temperature, depends on the magnetic configuration, as we will see). For ions in the $\sqrt{\nu}$ regime, and moderate values of $Z_I$, $(Z_I/Z_i)\delta_i - \delta_I$ is positive. Then the second term on the right-hand-side of equation~(\ref{EQ_SIMP}) is negative if $T_i$ is not hollow  (since $L_{11}^I$ is always positive~\cite{beidler2011ICNTS}), and a hollow impurity density profile cannot be sustained. In other words, for standard ion-root plasmas of flat density, $E_r \approx 5T_i'/4Z_ie$ (see equation~(\ref{EQ_ER})), which is always negative enough to produce impurity accumulation, since the corresponding thermodynamical force $5Z_IT_i'/4Z_iT_i$ always overcomes that of the impurity temperature gradient $\delta_IT_i'/T_i$. This is the argument usually recalled to explain the lack of temperature-screening in helical devices.

Nevertheless, this line of reasoning is not accurate in a plasma with very high ion and electron temperature, as it can be the case in large helical devices, since the collisionality ($\nu_b^*\!=\!R\nu/\iota v_{t_b}\!\propto\!n_b/T_b^2$, $\iota$ being the rotational transform, and $\nu_b$ and $v_{t_b}$ being the collision frequency and the thermal velocity of species $b$) of both species may become very small. For ions deep in the $\sqrt{\nu}$ regime and electrons in the $1/\nu$ regime, the assumption $L_{11}^e\ll L_{11}^i$ is not justified. If we return to equation~(\ref{EQ_FLUX}), we see that the third term in the right-hand-side is generally an outward pinch, but the sign of the second term is not defined. When electrons can be ignored, as in equation~(\ref{EQ_SIMP}), this term is directed inwards and, as we have seen, it is generally larger than the third one, leading to impurity accumulation. Nevertheless for very low collisionalities, with ion and electrons in the $\sqrt{\nu}$ and $1/\nu$ regimes respectively, we may have $\frac{7}{2}L_{11}^e\sim \frac{5}{4}L_{11}^i$, due precisely to the very different collisionality scaling of these two regimes. Altogether, due to very high $T_i$ and $T_e$, for moderate values of $Z_I$, one could be in a scenario such that the direction of the second term of the right-hand-side of equation~(\ref{EQ_FLUX}) is negative but small and thus smaller in absolute value than the third one, which is positive. If we look at equation~(\ref{EQ_ER}), this is equivalent to saying that the radial electric field is negative but small in absolute value, allowing for the transport associated to the gradient of $T_i$ to overcome the inwards transport associated to the negative $E_r$. This would mean an outward impurity pinch in a plasma with negative radial electric field. Of course, temperatures even higher will cause a positive radial electric field~\cite{yokoyama2002lhd} and an outward impurity flux. Plasmas of small and negative $E_r$ have been briefly discussed in predictive studies of bulk particle transport in Wendelstein 7-X (W7-X)~\cite{maassberg1999densitycontrol} and of energy transport in heliotron plasmas of high temperature~\cite{yokoyama2002lhd,yokoyama2010highti}; in this work we focus in its relevance for impurity transport.

\section{Parameter  dependence of the solution of small $E_r$}\label{SEC_SCA}

Since neoclassical transport depends on the magnetic configuration, we would like to see how the discussion of the previous section depends on the major radius of the device, its inverse aspect ratio $\epsilon$, and its average magnetic field $B$, as well as on the density and temperature of each species. The qualitative discussion of the previous section, although illustrative, is not accurate for our purposes. The reason is that, in the regimes of interest here, the $L_{11}^{i}(E_r)$ dependence cannot be ignored. Furthermore, we are interested in plasmas such that the inward pinch of high-$Z_I$ impurities due to the radial electric field is compensated by the outward pinch associated to the ion temperature gradient:
\begin{eqnarray}
|E_r| \sim \frac{\delta_I}{Z_Ie} |T_i'| \ll \frac{\delta_i}{Z_ie} |T_i'| \sim  \frac{T_i}{Z_ie\epsilon R}\,.\label{EQ_1SCA2}
\end{eqnarray}
In order to describe this situation, one can set to zero the left hand side of equation~(\ref{EQ_ER}) (this is equivalent to removing the pinch associated to the radial electric field in equation (\ref{EQ_AMB})) and see that the radial electric field is implicitly given~\cite{maassberg1999densitycontrol} by:
\begin{eqnarray}
\delta_eL_{11}^{e,1/\nu} = \delta_iL_{11}^{i,\sqrt{\nu}}(E_r)\,,\label{EQ_1SCA}
\end{eqnarray}
where we have assumed that the density profile is flat and that, due to the low collisionalities, the electrons and ions are in the $1/\nu$ and $\sqrt{\nu}$ regimes respectively.

The $1/\nu$ transport coefficient of species $b$ can be shown to scale as~\cite{hokulsrud1986neo,parra2016omni}:
\begin{eqnarray}
\delta_bL_{11}^{b,1/\nu} \sim \left(\frac{\nu_{b}R}{v_{t_b}}\right)^{-1}\left(\frac{\rho_b}{R}\right)^2\epsilon^{3/2}R v_{t_b}\,,
\end{eqnarray}
where $\rho_b$ is the Larmor radius of species $b$ and we have used that $\epsilon\!\sim\!a/R$, being $a$ the minor radius. These quantities are known to scale as:
\begin{eqnarray}
v_{t_b} &\sim & \left(\frac{T_b}{A_bm}\right)^{1/2}\,,\nonumber\\
\rho_b &\sim & \frac{(A_b m T_b)^{1/2}}{(Z_be)B}\,,\nonumber\\
\nu_{b} & \sim & \frac{n_bZ_b^4e^4}{\varepsilon_0^2(A_bm)^{1/2}T_b^{3/2}}\,,\label{EQ_NU}
\end{eqnarray}
where $m$ is the proton mass, $A_b\!\equiv\!m_b/m$ is the species mass number (in particular, $A_e\!\equiv\!m_e/m$), $\varepsilon_0$ the permittivity of free space, and for simplicity we do not consider inter-species collisions for the scaling of the flux (they are negligible for the ions and would introduce a $O(1)$ correction factor for the electrons in hydrogen or deuterium plasmas). For electrons, we end up with:
\begin{eqnarray}
\delta_eL_{11}^{e,1/\nu} &\sim & (A_e^{1/2}m^{1/2}e^{-6}\varepsilon_0^2)(\epsilon^{3/2}R^{-2}B^{-2})(n_e^{-1}T_e^{7/2})\,.\label{EQ_1NU}
\end{eqnarray}
The $\sqrt{\nu}$ particle flux can be shown to scale as~\cite{hokulsrud1986neo,calvo2016sqrtnu}:
\begin{eqnarray}
\delta_bL_{11}^{b,\sqrt{\nu}} \sim \frac{T_b^{1/2}}{(A_bm)^{1/2}\epsilon^{3/2}R}\left(\frac{\nu_{b}}{\omega_\alpha^{3}}\right)^{1/2} \left(\frac{\rho_b}{R}\right)^2\epsilon^{3/2} R v_{t_b}\,,
\end{eqnarray}
where $\omega_\alpha$ is the precession frequency (associated to the motion caused by the drifts that are tangential to the flux surface). In cases where the tangential magnetic drift can be neglected, the latter scales with the radial electric field as:
\begin{eqnarray}
\omega_\alpha\sim \frac{E_r}{\epsilon RB}\,.\label{EQ_EXB}
\end{eqnarray}
In general cases, one can use that $Z_seE_r\sim T_s /\epsilon R$ if $s$ is the so-called rate-controlling species (i.e., $s\!=\!e$ in electron root and $s\!=\!i$ in ion root), but in this work we are interested in allowing for the radial electric field to be smaller (i.e., to follow equation~(\ref{EQ_1SCA2})). In the trace-impurity limit, we end up with:
\begin{eqnarray}
\delta_iL_{11}^{i,\sqrt{\nu}} &\sim & (m^{-1/4}\varepsilon_0^{-1})(\epsilon^{3/2} R^{-1/2}B^{-1/2})(A_i^{-1/4}Z^{-1/2})(n_e^{1/2}T_i^{5/4}E_r^{-3/2})\,.\label{EQ_SQRTNU}
\end{eqnarray}
where we have dropped $\delta_e$ and $\delta_i$, since they are $O(1)$ factors. If we now combine equations~(\ref{EQ_1SCA}), (\ref{EQ_1NU}), and (\ref{EQ_SQRTNU}), we obtain:
\begin{eqnarray}
E_r \sim (A_e^{-1/3}m^{-1/2}e^{4}\varepsilon_0^{-2})(RB)(A_i^{-1/6}Z_i^{-1/3})(n_e T_i^{5/6}T_e^{-7/3})\,,
\end{eqnarray}
which for thermally coupled species ($T_e\!=\!T_i$) becomes:
\begin{eqnarray}
E_r \sim (A_e^{-1/3}m^{-1/2}e^{4}\varepsilon_0^{-2})(RB)(A_i^{-1/6}Z_i^{-1/3})(n_e T_i^{-3/2})\,.\label{EQ_FSER}
\end{eqnarray}

The scaling of equation (\ref{EQ_FSER}) predicts that the radial electric field becomes smaller in absolute value as $n_e T_i^{-3/2}$, a quantity close to the collisionality, becomes small, as discussed in the previous section. 

Before studying its effect on impurity transport, let us asses the range of validity of the scaling of equation~(\ref{EQ_FSER}). First, we must impose that the plasmas have indeed low collisionality, below the plateau {(for which $L_{11}^e \ll L_{11}^i$)}:
\begin{eqnarray}
\nu_b^* &\ll& \epsilon^{3/2}\,,\label{EQ_PLATEAU}
\end{eqnarray}
for both electrons and ions. Once in the low-collisionality regime, the relation between the $E\times B$ rotation frequency and the collision frequency must be such that the ions are in the $\sqrt{\nu}$ regime and the electrons are in the $1/\nu$ regime. It can be shown~\cite{calvo2016sqrtnu} that this happens when:
\begin{eqnarray}
L_{11}^{e,1/\nu} &\ll& L_{11}^{e,\sqrt{\nu}}\,,\nonumber\\
L_{11}^{i,1/\nu} &\gg& L_{11}^{i,\sqrt{\nu}}\,,
\end{eqnarray}
which can be written as:
\begin{eqnarray}
\epsilon^{3/2} \left(\frac{E_r}{\epsilon R B\nu_i}\right)^{3/2} &\gg& 1\,,\label{EQ_1NUSQRTNU1}\\
\epsilon^{3/2} \left(\frac{E_r}{\epsilon R B\nu_e}\right)^{3/2} &\ll& 1\,.\label{EQ_1NUSQRTNU2}
\end{eqnarray}
One can show that equation (\ref{EQ_1SCA}) can be recasted as:
\begin{equation}
\epsilon^{3/2} \left(\frac{E_r}{\epsilon R B}\right)^{3/2}\frac{1}{\nu_e\nu_i^{1/2}} \sim 1\,.
\end{equation}
It is now easy to see that the conditions of equations (\ref{EQ_1NUSQRTNU1}) and (\ref{EQ_1NUSQRTNU2}) are equivalent to:
\begin{eqnarray}
 \nu_i &\ll& \nu_e\,,\nonumber\\
 \nu_i^{1/2} &\ll& \nu_e^{1/2}\,,
\end{eqnarray}
respectively. The latter is more restrictive, and can be explicitly written as:
\begin{eqnarray}
Z_i^{3/2}A_i^{-1/4}\ll A_e^{-1/4} \approx 6.5\,,
\end{eqnarray}
which is automatically fulfilled for hydrogen or deuterium plasmas. For small enough radial electric field, the magnetic drift tangential to the flux surface may become relevant and ions and electrons may enter the superbanana-plateau regime, in which the fluxes do not scale with the collision frequency (see e.g.~\cite{calvo2016sqrtnu}). Equilibration of the magnetic and $E\times B$ drifts gives:
\begin{eqnarray}
\omega_\alpha\sim \frac{\rho_b}{R}\frac{v_{t_b}}{\epsilon R} \sim \frac{T_b}{Z_be \epsilon R^2 B}\,.\label{EQ_VMPOL}
\end{eqnarray}
Comparing equation (\ref{EQ_VMPOL}) with equation (\ref{EQ_EXB}), we see that this situation is avoided if:
\begin{eqnarray}
E_r \gg \frac{T_b}{Z_be R}\,,\label{EQ_SBP}
\end{eqnarray}
which sets a lower limit to the size of the radial electric field. Finally, for low enough collisionalities, the ions could be in the $\nu$ regime. We will not worry about this possibility, since the consequences of this change of regime for our scaling would be relatively small (as $L_{11}^{i,{\nu}}\!\sim T_i^{1/2}E_r^{-2}$~\cite{beidler2011ICNTS}, to be compared to $L_{11}^{i,\sqrt{\nu}}\!\sim T_i^{5/4}E_r^{-3/2}$). 

\begin{figure}
\begin{center}
\includegraphics[angle=0,width=\columnwidth]{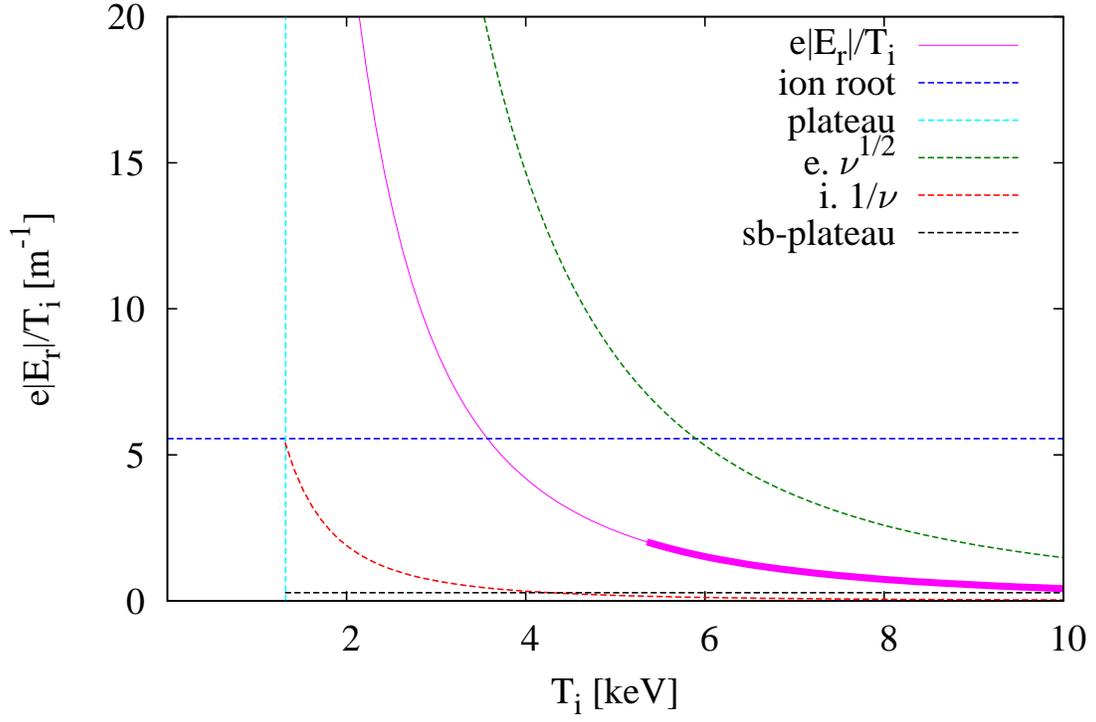}\vskip-1.5cm
\includegraphics[angle=0,width=\columnwidth]{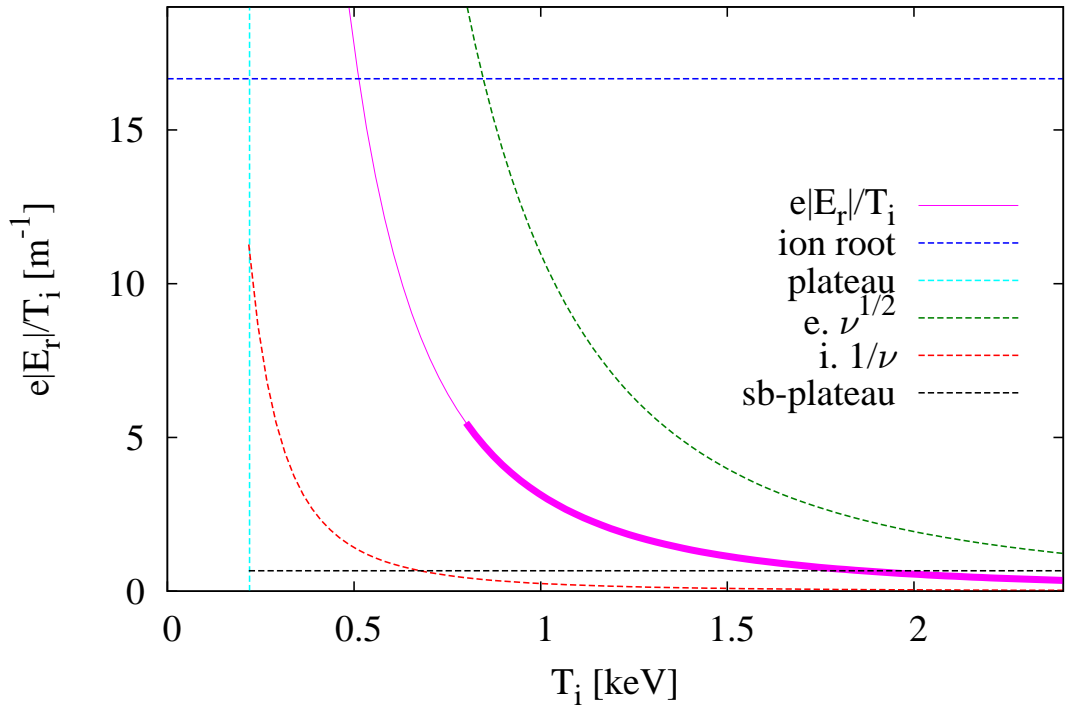}\vskip-1.5cm
\end{center}
\caption{Diagram of regimes for an LHD-size device (top) and a TJ-II-size device (bottom). Detailed description in the text.}\label{FIG_DIA}
\end{figure}

Figure~(\ref{FIG_DIA}) (top) shows schematically the derived conditions in a diagram of $eE_r/T_i$ versus $T_i$, for an LHD-size device ($R\!=\!3.5\,$m, $\epsilon\!=\!0.05$, $B\!=\!3\,$T) for $n_e\!=\!2\times 10^{19}\,$m$^{-3}$. In figure~(\ref{FIG_DIA}) (bottom) we show the same diagram for a TJ-II-size device ($R\!=\!1.5\,$m, $\epsilon\!=\!0.04$, $B\!=\!1\,$T) for $n_e\!=\!0.3\times 10^{19}\,$m$^{-3}$. In both cases, we set $Z_i\!=\!1$, $A_i\!=\!1$ and $T_e\!=\!T_i$. Equation~(\ref{EQ_FSER}) is plotted in magenta: the thick part of the line corresponds to the parameter space that we are interested in; temperatures left of the cyan line are discarded since they correspond to the plateau regime (i.e., they do not fulfill equation~(\ref{EQ_PLATEAU})); values of $e|E_r|/T_i$ above {or slightly below} the blue line correspond to standard ion root plasmas {($e|E_r|/T_i=1/(Z_ie\epsilon R)$ for the blue line and $e|E_r|/T_i<1/(3Z_ie\epsilon R)$ for the thick magenta line)}, and below the black line to plasmas with contribution of the superbanana-plateau regime (since they violate the conditions of equations (\ref{EQ_1SCA2}) and (\ref{EQ_SBP}) respectively); in the top right part of the plot delimited by the green line, the electrons are in the $\sqrt{\nu}$ regime (see equation (\ref{EQ_1NUSQRTNU1})); in the bottom left part of the plot, enclosed by the red, cyan and black lines, there is a region with the ions in the $1/\nu$ regime (equation (\ref{EQ_1NUSQRTNU2})).

Before discussing the specific details of each figure, we must note that the above scalings neglect $O(1)$ factors {and dimensionless factors related to the optimization, see below; this may cause the scalings deviate from accurate calculations (which we will show in following sections), so attention should be paid to the relative position of the curves rather than to the absolute numbers}. With this caveat in mind, we see that there may exist a wide range of temperatures in which in a small enough radial electric field can be realized: the plateau collisionality is well below the temperatures required for {$L_{11}^e \sim L_{11}^i$ (i.e., the cyan line always falls to the left of the magenta line)}. As we have discussed, as the collisionality decreases, so does the radial electric field, and the electrons never go into the $\sqrt{\nu}$ regime (the magenta line always lies between the green and red lines), unless an additional electron root appears. The superbanana-plateau regime might become relevant at very high temperatures. If this were the case, the lack of scaling with the collision frequency, and the mass difference between electrons and ion would be inconsistent with equation (\ref{EQ_1SCA}), and $E_r$ would tend to become more negative. Nevertheless, the experimental evidence available suggests that this situation will not take place: at high enough temperature, the additional positive electron root that is expected to appear~\cite{yokoyama2002lhd} has been observed in the experiment~\cite{nagaoka2015etb}.

If we now combine equations~(\ref{EQ_1SCA2}) and (\ref{EQ_FSER}) we can estimate the temperature at which the pinches balance each other:
\begin{eqnarray}
n_e T_i^{-5/2} \sim (A_e^{1/3}m^{1/2}e^{-5}\varepsilon_0^2)(R^{-2}B^{-1})(A_i^{1/6}Z_i^{-2/3})Z_I^{-1}\,.\label{EQ_FSCA}
\end{eqnarray}
Equation~(\ref{EQ_FSCA}) shows that $n_e T_i^{-5/2}$, again a quantity similar to the collisionality, is the relevant quantity: for a given magnetic configuration and working gas, it determines whether we are in a situation where the radial electric field can be much smaller than the ion root prediction. From equation (\ref{EQ_FSCA}) and figure (\ref{FIG_DIA}) we learn that for large devices low collisionalities are needed (note the negative exponent in $R$ and $B$); for small devices, this can be achieved at higher collisionalities. We will test these scalings with simulations and experiments for LHD and TJ-II.

{We finally note that equations~(\ref{EQ_NU}) and (\ref{EQ_SQRTNU}) do not include the effect of optimization of the magnetic configuration with respect to neoclassical transport. The degree of optimization can be described with $\delta$, the size of the deviation of the magnetic field ($B_{omni} + \delta B_{dev}$, with $B_{omni}\sim B_{dev}$) from perfect omnigeneity ($B_{omni}$) see~\cite{parra2016omni,calvo2016sqrtnu} for details (and note that $\delta$ is related to the so-called effective ripple~\cite{beidler2011ICNTS}). It can be shown that both expressions should be multiplied by some power of $\delta$, and that the exponent may take different values depending on the helicities present in $B_{dev}$. Since a universal scaling is not available, in this paper we will limit ourselves to study the effect of optimization within the parameter space of the LHD device.}

\begin{figure}
\begin{center}
\includegraphics[angle=0,width=\columnwidth]{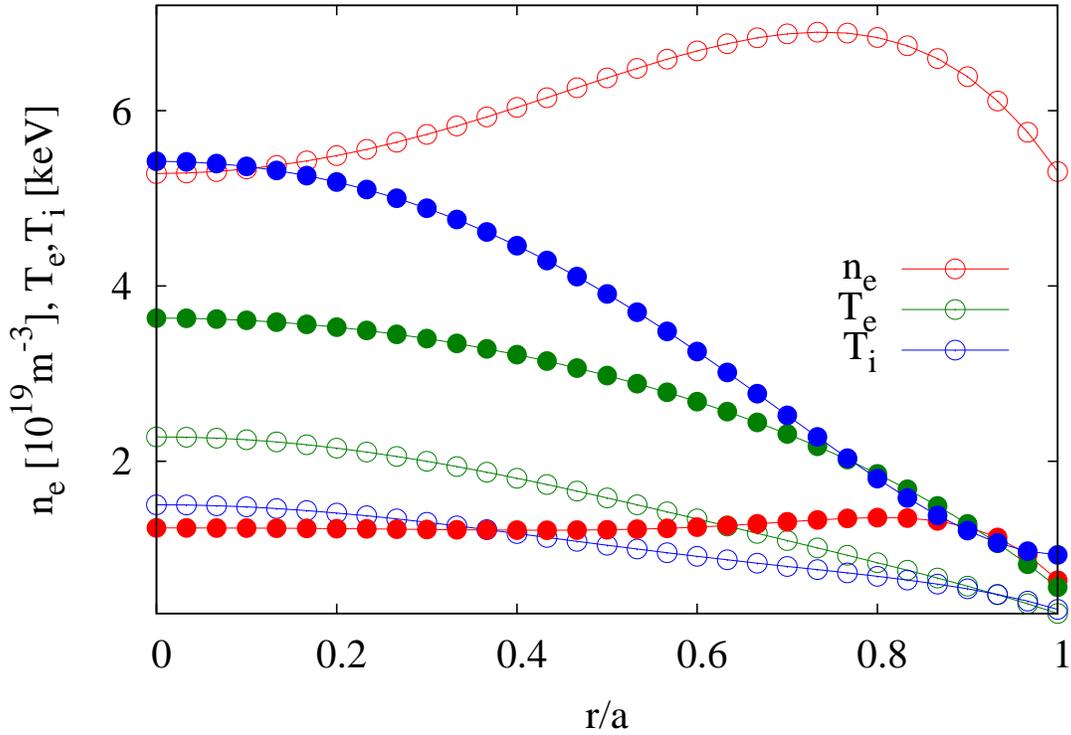}\vskip-1.5cm
\end{center}
\caption{Density and temperature profiles of LHD discharges \#109696, at $t\!=\!4.440\,$s (open circles) and \#113208 at $t\!=\!4.640\,$s (closed circles).}\label{FIG_PLASMAS_LHD}
\end{figure}

\begin{figure}
\begin{center}
\includegraphics[angle=0,width=\columnwidth]{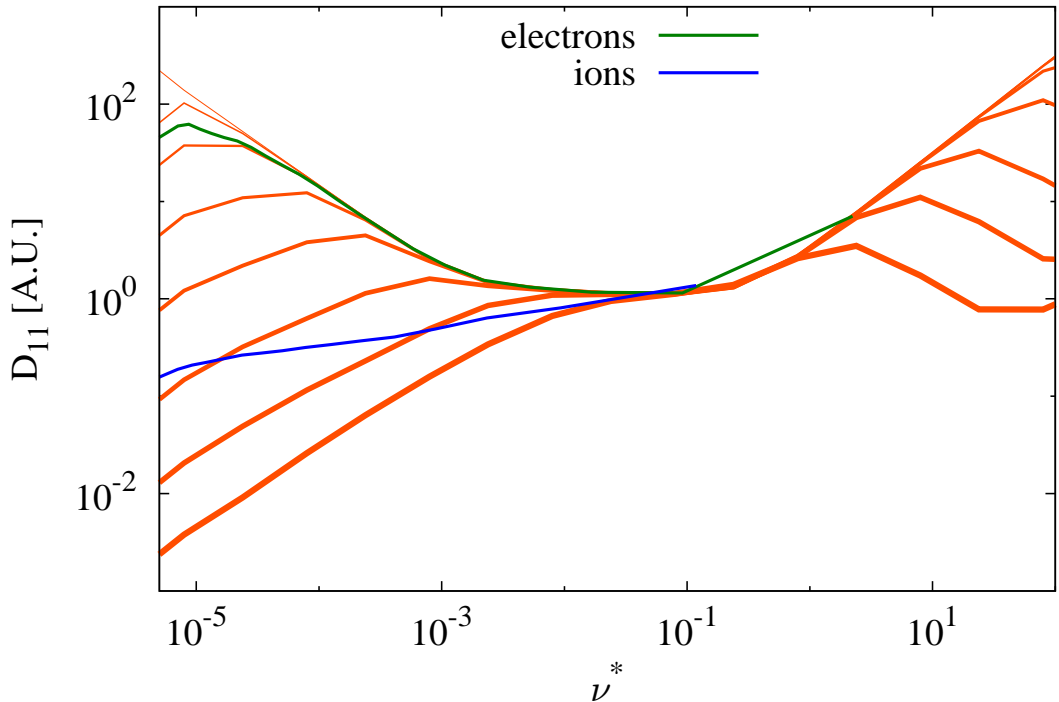}\vskip-1cm
\end{center}
\caption{{Monoenergetic transport coefficient $D_{11}$ (see~\cite{beidler2011ICNTS}) at $r/a\!=\!0.3$. Different orange lines correspond to different values of $|E_r|/v$, $v$ being the particle velocity, and thicker lines indicating larger $|E_r|/v$; {for the blue and green curves, $E_r$ is held fixed and equal to that given by ambipolarity of the neoclassical fluxes.}}}\label{FIG_MONO}
\end{figure}

\begin{figure}
\begin{center}
\includegraphics[angle=0,width=\columnwidth]{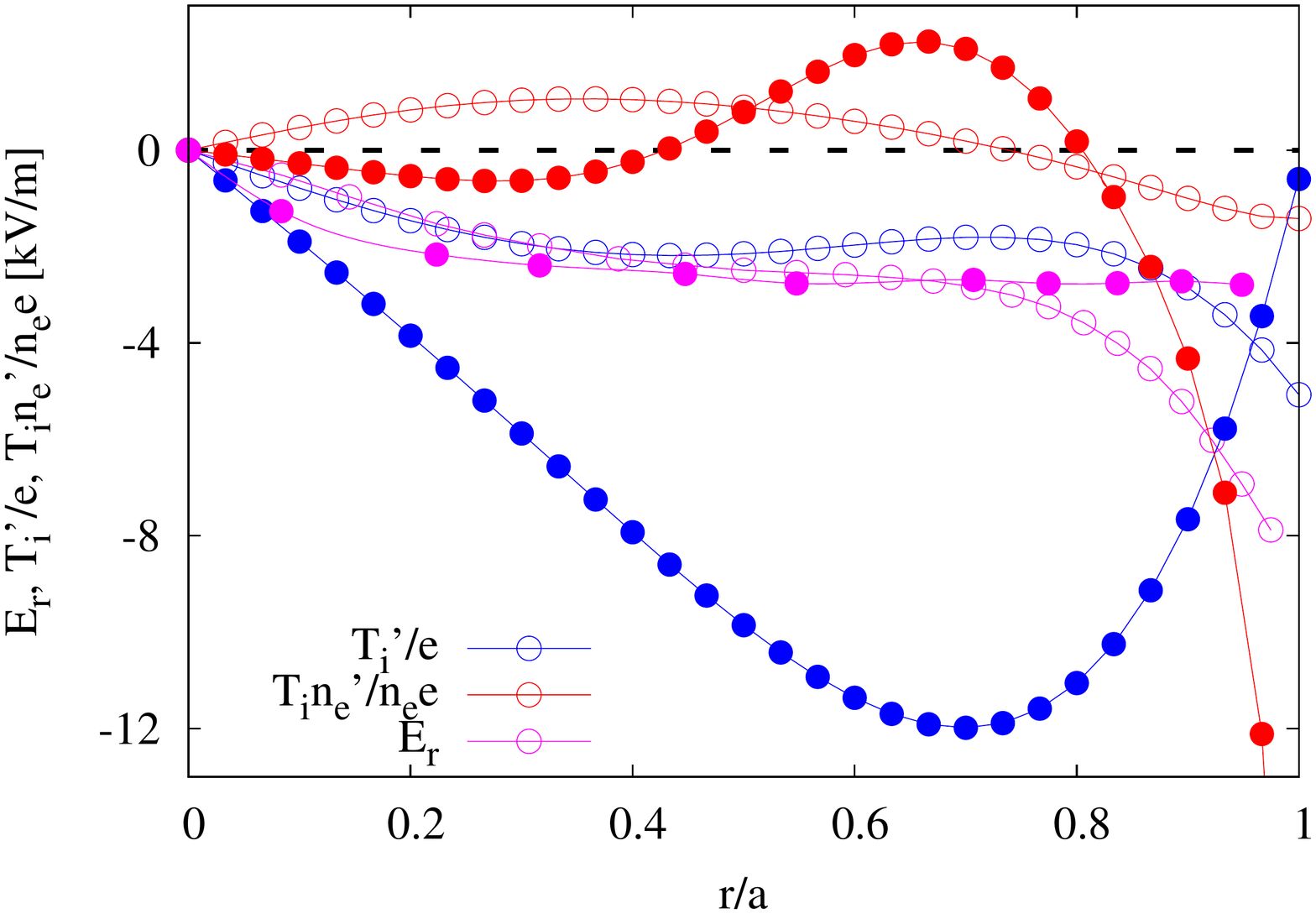}\vskip-1.5cm
\end{center}
\caption{Ion temperature gradient and calculated radial electric field of LHD discharges {\#109696}, at $t\!=\!4.440\,$s (open circles) and \#113208 at $t\!=\!4.640\,$s (closed circles).}\label{FIG_RESULTS_LHD}
\end{figure}

\section{LHD}\label{SEC_LHD}

Let us calculate the relevant terms in equation~(\ref{EQ_FLUX}) for two real discharges in the inward-shifted magnetic configuration of LHD. The first one (\#109696, $t\!=\!4.440\,$s) is a medium-density LHD hydrogen plasma with highest input power; this discharge has been documented in~\cite{dinklage2013ncval,satake2015ishw} as part of an intermachine validation of neoclassical transport predictions for ion-root plasmas, and it is available at the ISHPDB. The second shot (\#113208, $t\!=\!4.640\,$s) is an impurity hole scenario as described in the introduction, with a mix of hydrogen, helium and carbon; first neoclassical multispecies simulations of this plasma will be reported in~\cite{sato2016penta} (using PENTA code~\cite{spong2005flow} based on the momentum-correction method~\cite{sugama2002viscosity}), and predictions of impurity transport (neoclassical and gyrokinetic) will be presented in~\cite{nunami2016iaea} (using the electromagnetic multispecies GKV code~\cite{ishizawa2015GKV,nunami2015GKV}); in this work we assume, for the sake of simplicity, a pure hydrogen plasma. 

The plasma profiles are shown together in figure~\ref{FIG_PLASMAS_LHD}. Both density profiles are hollow, with core density about 5 times larger in the ion root plasma; the temperatures are peaked, larger in the impurity hole plasma (with $T_i\!>\!T_e$) than in the ion root plasma (where $T_i\!<\!T_e$). As a consequence of this, the collisionality in the core of the impurity hole plasma is $\sim$ 20 times lower than for the standard ion-root plasma: regarding the ions, the former is deep in the $\sqrt{\nu}$ regime, while the latter is close to the plateau boundary~\cite{satake2015ishw}. In both cases the collisionality of the electrons is relatively close to that of the ions, when compared with core electron root confinement (CERC) plasmas ~\cite{yokoyama2007cerc}, in which $\nu_e^*\!\ll\!\nu_i^*$. 

We now calculate solutions for LHD of the equations discussed in this and the previous section using the Drift Kinetic Equation Solver (DKES)~\cite{hirshman1986dkes}: details of the calculation and convolution of the monoenergetic coefficients including momentum-correction techniques~\cite{maassberg2009momentum} may be found in~\cite{velasco2011bootstrap}. DKES solves the monoenergetic drift kinetic equation (DKE), which means a series of simplifications with respect to the complete DKE. This might affect some of the quantitative predictions~\cite{satake2006fortec3d,landreman2014sfincs,satake2006fortec3d,satake2015ishw,velasco2016pellets}, but not the qualitative discussion, which is based on general scalings that remain valid when the full DKE is solved.

We first show schematically in figure~\ref{FIG_MONO} the monoenergetic transport coefficients~\cite{beidler2011ICNTS} calculated with DKES at $r/a\!=\!0.3$, where it is apparent that the electrons in the impurity hole plasma are mainly in the $1/\nu$ regime and the ions in the $\sqrt{\nu}$ regimes respectively: for {the range of collisionalities relevant for particles following a Maxwellian distribution with the density and temperatures of discharge \#113208}, the monoenergetic diffusion coefficient of electrons (green) scales with the inverse of the collisionality (with the exception of the electrons in the tail of the Maxwellian, in the left part of the figure), while for the ions (blue), due to its smaller thermal velocity, the large value of $|E_r|/v$ removes this dependence. As we have discussed, the scaling of section~\ref{SEC_SCA} is valid as long as each species remains in the corresponding collisionality regime. For very low collisionalities, the ions may be in the $\nu$ regime. In such situation, reducing the collisionality would still bring the size of the electron flux closer to that of the ions, but more slowly, since the difference in collisionality dependence is smaller. DKES does not include the magnetic drift within the flux surface, but we do not expect the superbanana-plateau regime to be relevant in these plasmas.

In figure~\ref{FIG_RESULTS_LHD} we show the results of neoclassical simulations of equation~(\ref{EQ_AMB}). In both cases the radial electric field is negative, in agreement with the experiment: in the ion root plasma, the neoclassical estimate was shown to slightly underestimate the absolute value of $E_r$ measured with charge-exchange recombination spectroscopy (CXRS)~\cite{dinklage2013ncval}; for impurity hole plasmas, values around $-3\,$kV/m have been measured with Heavy Ion Beam Probe~\cite{ido2010hibp}, in agreement with the simulations presented in this work. If we now return to the discussion of the previous section, we see that for the ion root plasma $E_r \approx 5T_i'/4Z_ie$ in a broad radial range, showing that the ambipolar equation can be described with ions in the $\sqrt{\nu}$ regime and neglecting the electron contribution; the familiar discussion on the lack of temperature screening holds. Nevertheless, for the impurity hole plasma, the radial electric field is several times smaller in absolute value than the standard ion-root prediction. This means that, for moderate values of the impurity charge number ($Z_I\!\approx 5$), the pinch term associated to the ion temperature gradient can compensate or overcome the radial electric field and produce an outward pinch. This can be the case for helium ash ($Z_I\!=\!2$) and some charge states of carbon. In both cases, the hollowness of the electron density profile tends to make the radial electric field less negative in some radial regions, as observed in~\cite{dinklage2016iaea} in experiments with pellets, and would tend to reduce the impurity influx; nevertheless, for the plasma conditions of impurity hole, the thermodynamical force associated to the density gradient is negligible when compared to that of the ion temperature gradient.

This compensation between the two pinch terms depends critically on details of the plasma profiles and on the impurity charge number. This criticality, one the one hand, would explain how a neoclassical mechanism (usually a smooth function on the plasma profiles) can contribute to a physical phenomena experimentally shown to depend very abruptly on the plasma parameters~\cite{yoshinuma2015reversal}. We can see this by imposing the zero-flux condition, $\Gamma_I\!=\!0$, in equation (\ref{EQ_SIMP}), which yields $n_I\!\sim\!T_i^{(Z_I/Z_i)\delta_i - \delta_I}$. Although we do not expect this condition to be reliable for describing the radial distribution of impurities, it is useful for illustrating that a change in the sign of ${(Z_I/Z_i)\delta_i - \delta_I}$ may cause a qualitative change in the impurity density. 

\begin{figure}
\begin{center}\vskip-2cm
\includegraphics[angle=0,width=\columnwidth]{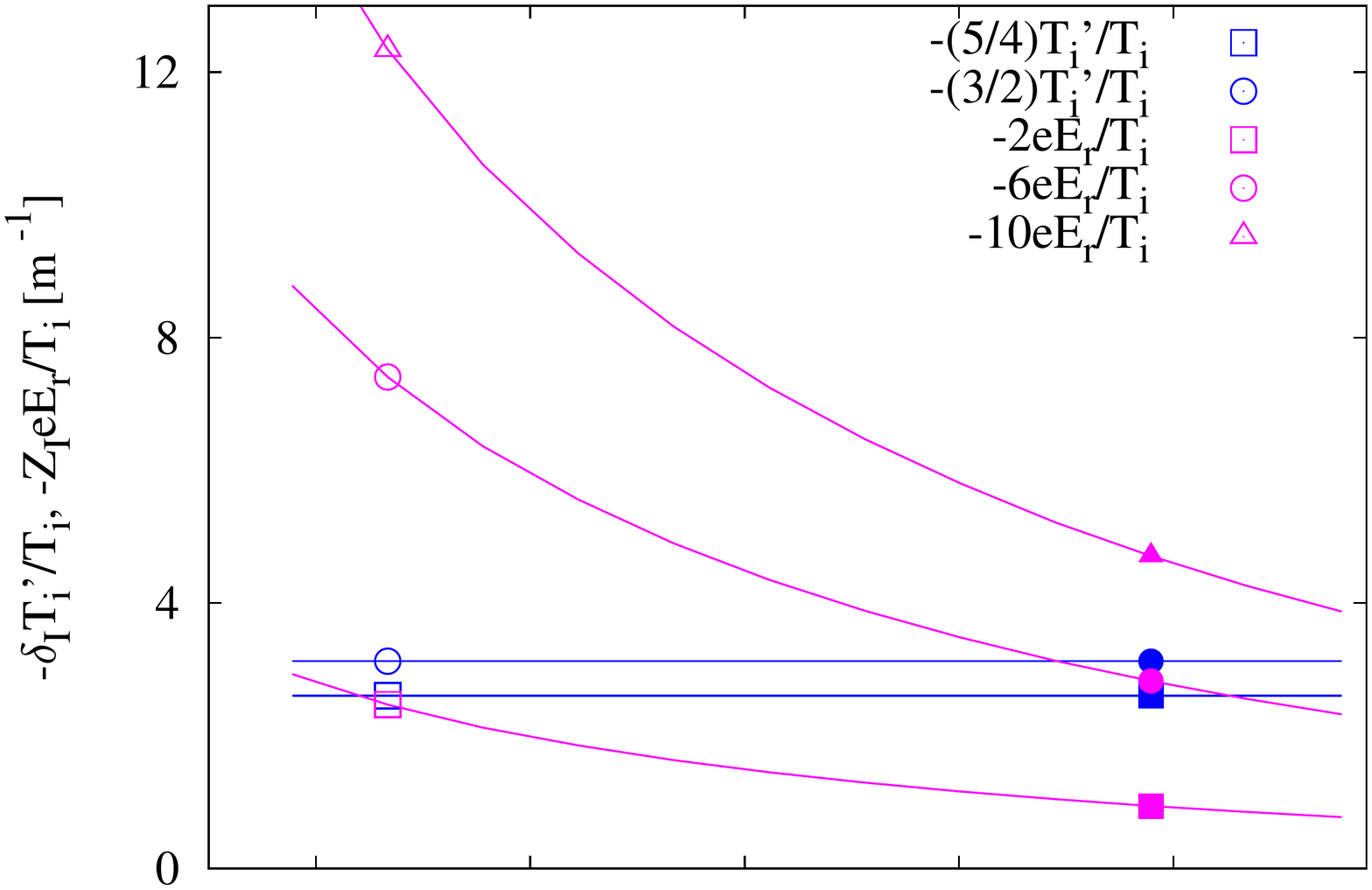}\vskip-3.5cm
\includegraphics[angle=0,width=\columnwidth]{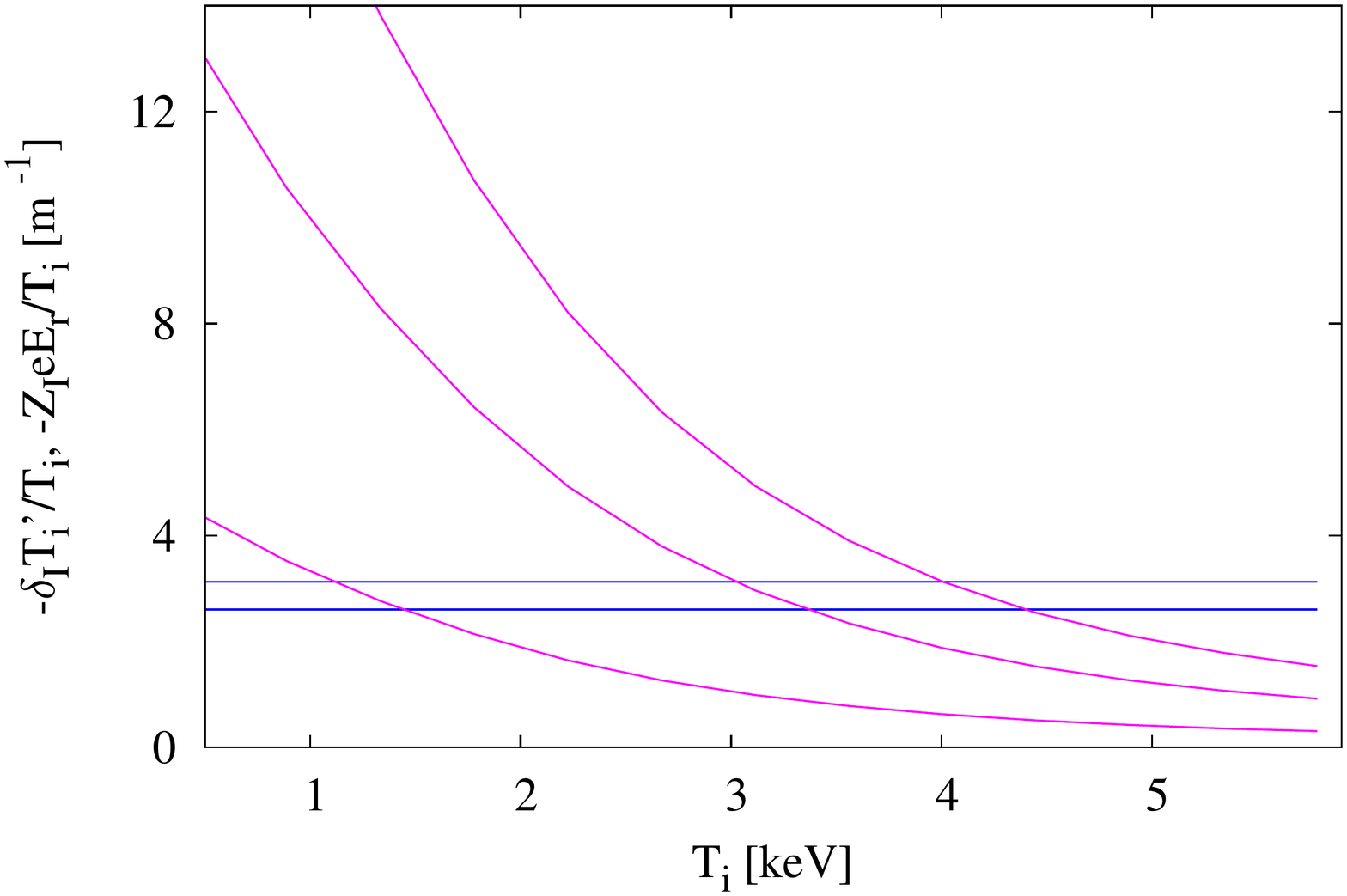}\vskip-2.5cm
\end{center}
\caption{{Convective terms in the impurity flux as a function of the ion temperature. Open and closed signs \changes{denote temperatures corresponding to the ion root and impurity hole plasma respectively. \changes{Scan with $T_i\!=\!2T_e\!-\!1\,$keV} (top) and $T_i\!=\!T_e$ (bottom).\label{FIG_SCAN}}}}
\end{figure}

On the other hand, although abrupt, the impurity hole has been consistently observed for several species and different heating schemes and configurations; therefore a robust mechanism has to be invoked for discussing it. In order to show the relevance of the combination of very high ion and electron temperatures we focus on radial position $r/a\!=\!0.3$. We then scan in ion and electron temperature, with $T_i$ from 1 to 5$\,$keV and $T_e$ from 1 to 3$\,$keV in the magnetic axis, \changes{while keeping $n_e$, $T_i'/T_i$ and $T_e'/T_e$ constant (and equal to those of the impurity hole discharge) during the scan}~\cite{ida2009observation}, and solve the ambipolar equation in the trace impurity limit. This is a similar exercise to the one of figure~\ref{FIG_DIA}, but now solving the ambipolar equation with DKES, without making any assumption on the collisionality regime of each species. The two LHD discharges analyzed in this work are marked with points in figure~\ref{FIG_SCAN} (similar scans were performed in~\cite{yokoyama2002lhd,yokoyama2010highti}, focusing on energy transport and using extrapolated scenarios). In figure~\ref{FIG_SCAN} (top) we show how the pinches associated to the radial electric field and the temperature gradient change, for several values of $Z_I$ and for impurities in two regimes: $\sqrt{\nu}$ ($\delta_I\!=\!5/4$) and plateau ($\delta_I\!=\!3/2$). The tendency predicted in Section~\ref{SEC_SCA} is clearly observed: as the temperature increases, the pinch associated to the radial electric field evolves from being much larger than that of the temperature gradient (standard ion-root operation) to be comparable to it (impurity hole type plasmas).

In the previous scan, the ion temperature went from being equal to the electron temperature to be about 50\% larger than it. In order to show that there is no qualitative change if the bulk species are more tightly thermally coupled, we repeat the scan with $T_i\!=\!T_e$ from 1 to 4$\,$keV in the magnetic axis, again keeping $T_i'/T_i$ constant. The results are shown in figure~\ref{FIG_SCAN} (bottom). The value of $T_i$ at which the total pinch is positive for a given $Z_I$ and impurity regime changes, but the parameter dependence is the same.

Figure~\ref{FIG_SCAN} indicates that, if one further increases the temperature, the radial electric field should become positive. Indeed, one would have $\frac{7}{2}L_{11}^e\gg \frac{5}{4}L_{11}^i$ in equation~(\ref{EQ_ER}), and both terms in the left-hand-side would become positive, as previously discussed~\cite{yokoyama2002lhd,nagaoka2015etb}.

 At this point, it is important to emphasize that impurity hole formation cannot be explained only based on the mechanisms discussed above: for real plasma conditions and relevant impurities, total inwards impurity pinch is consistently predicted by neoclassical simulations~\cite{ido2010hibp,regana2013euterpe}. What our simulations show is that plasmas of very low ion collisionality are a scenario in which the standard neoclassical pinch of impurities (caused by $T_i'$ and $E_r$) can be relatively small in absolute value, when compared with standard ion-root plasmas. This opens the possibility that the small inward neoclassical pinch is overcome by other terms (associated e.g. to asymmetries or to turbulence) that otherwise are negligible. These mechanisms are probably required for explaining the charge dependence of the impurity hole: in the experiment, at a given temperature, impurity hole is more prominent for higher $Z_I$; nevertheless, according to our neoclassical predictions, higher $T_i$ should be needed for higher values of $Z_I$ (these impurities could have larger $\delta_I$, but it is unlikely that large enough to compensate the much larger inwards $Z_IE_r$ pinch). Only if the radial electric field were positive, the charge dependence of the impurity hole would be straightforwardly understandable, but this is in contradiction with available measurements. Figure~\ref{FIG_SCAN} shows that, for low enough ion collisionality, the neoclassical pinch will always be directed outwards; nevertheless, the reduction of high-$Z_I$ impurity content is predicted to take place at values of $T_i$ higher than those seen in the experiment.

\begin{figure}
\begin{center}
\includegraphics[angle=0,width=0.7\columnwidth]{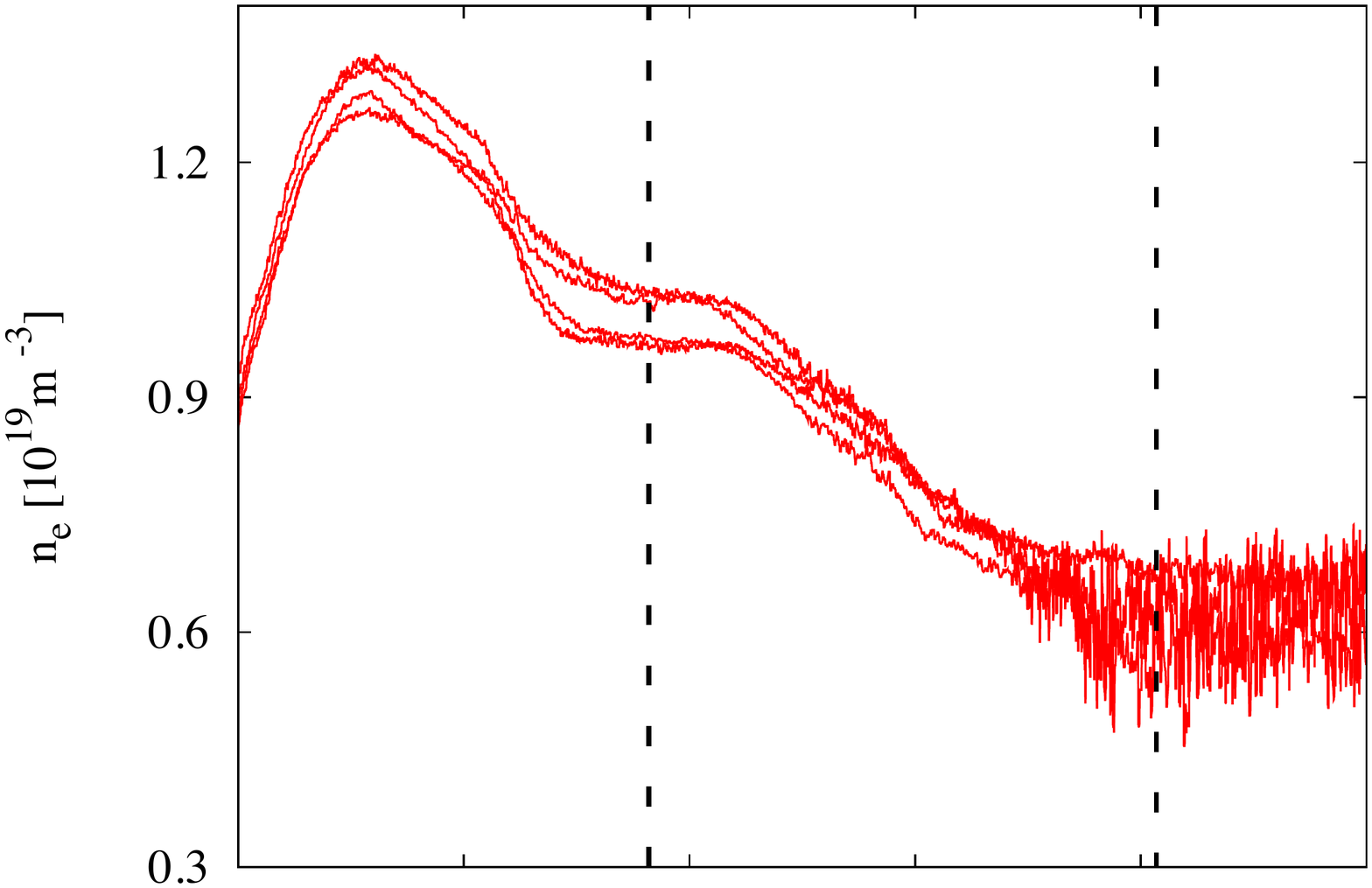}\vskip-2.5cm
\includegraphics[angle=0,width=0.7\columnwidth]{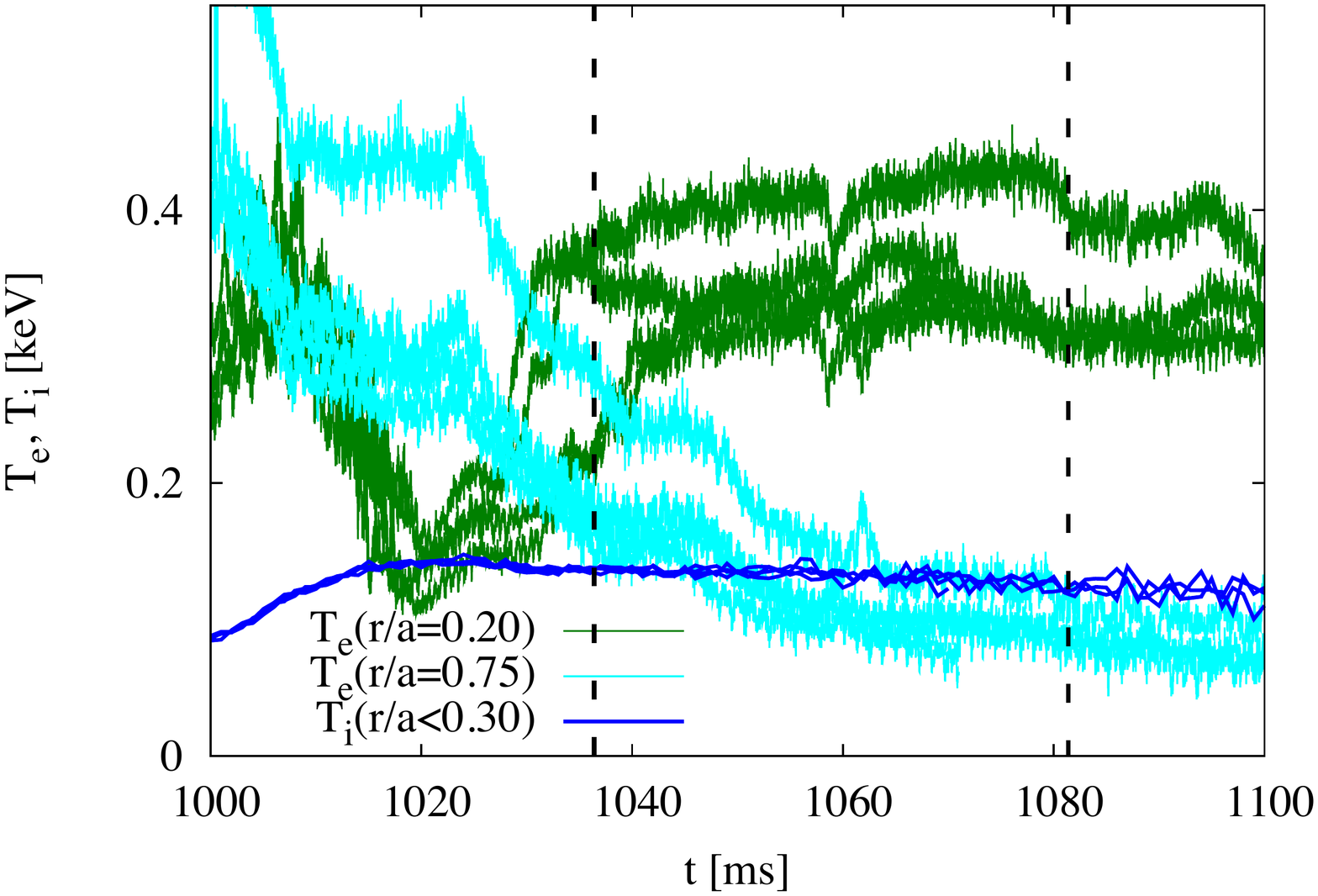}\vskip-2.5cm
\end{center}
\caption{{Time traces for TJ-II discharges \#41706, \#41709-41711: (top) line-averaged density, (bottom) central and edge electron temperature and central ion temperature.}}\label{FIG_TRACES_TJ-II}
\end{figure}

As for the configuration dependence within a single device, it cannot be discussed with the simple scaling of equation~(\ref{EQ_FSCA}), and neoclassical optimization must be taken into account. Neoclassical simulations have shown that the outward-shifted magnetic configuration of LHD has a larger $1/\nu$ transport with respect to the optimized inward-shifted configuration, while the increase in the $\sqrt{\nu}$ transport is smaller (see e.g. a comparison in~\cite{beidler2011ICNTS}), which makes easier the transition to CERC~\cite{yokoyama2002lhd,ida2005itb}. Within our framework, for the same reason, better impurity behaviour would be expected in the outward-shifted configuration, as observed in the experiment~\cite{ida2009observation,yoshinuma2009observation}.
 
\begin{figure}
\begin{center}
\includegraphics[angle=0,width=0.7\columnwidth]{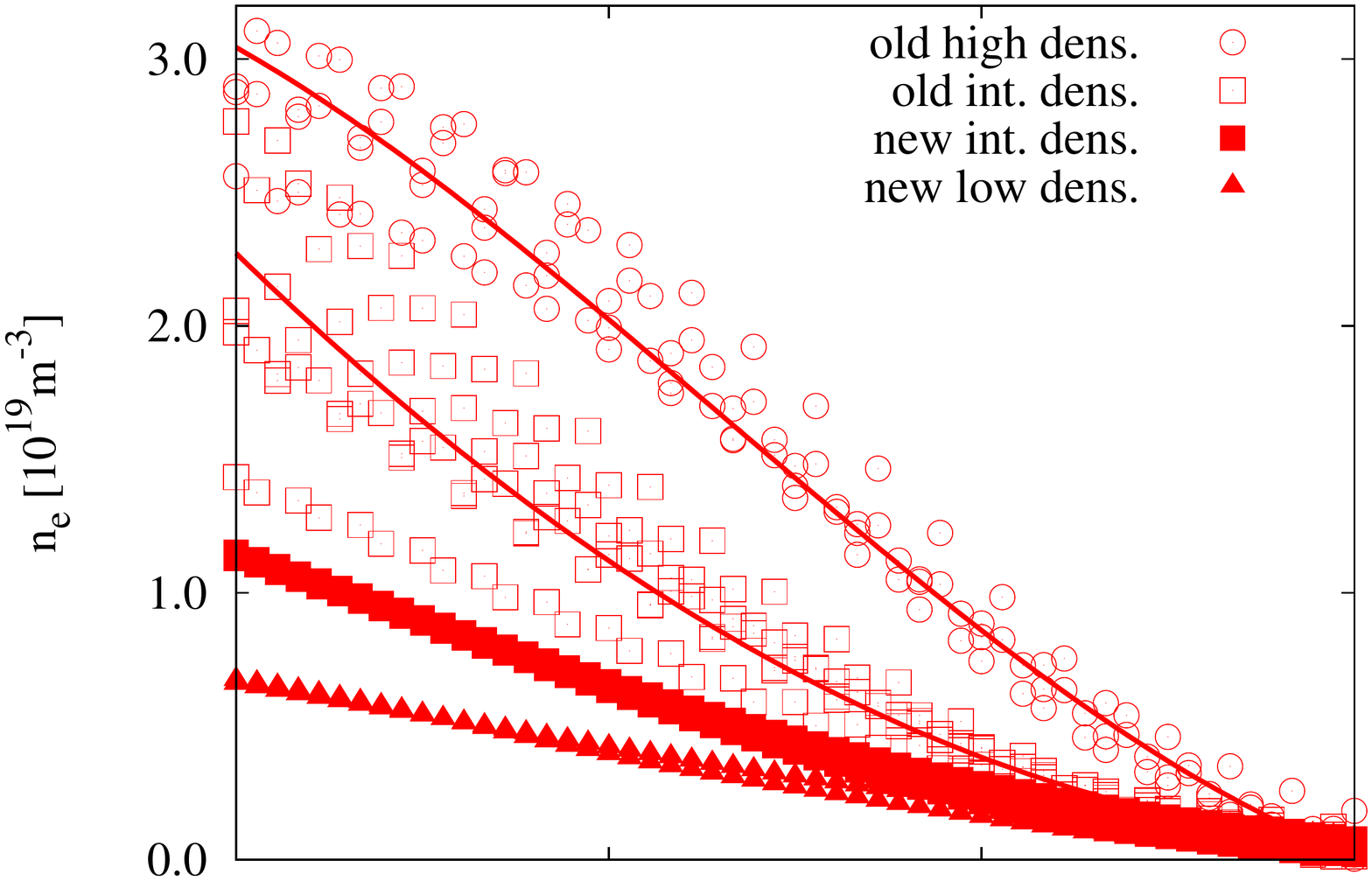}\vskip-2.5cm
\includegraphics[angle=0,width=0.7\columnwidth]{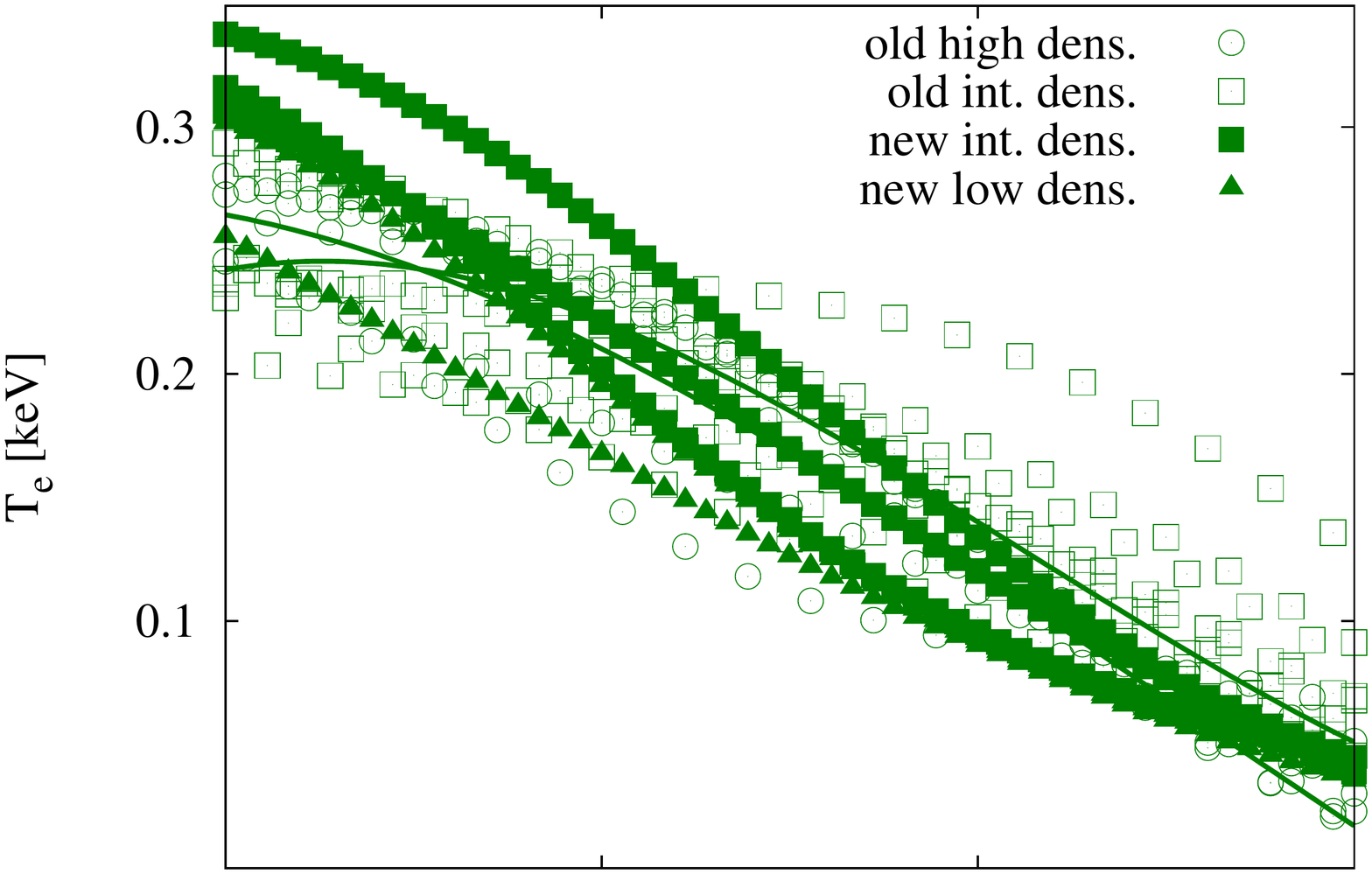}\vskip-2.5cm
\includegraphics[angle=0,width=0.7\columnwidth]{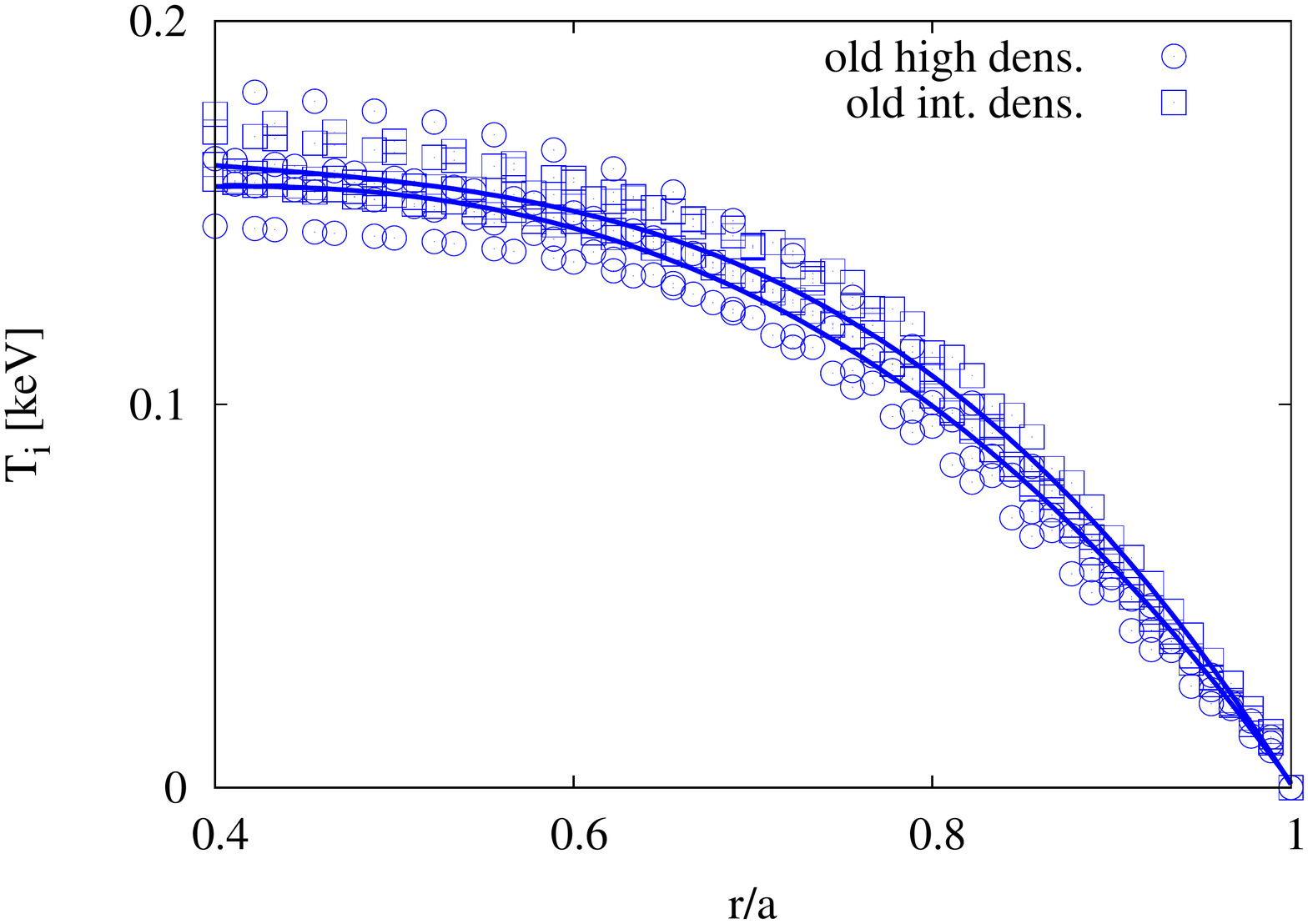}\vskip-2.5cm
\end{center}
\caption{{Density and temperature profiles of TJ-II discharges for high density (open circles), intermediate density (open squares), new discharges with intermediate density (closed squares) and low density (closed triangles).}}\label{FIG_PLASMAS_TJ-II}
\end{figure}

\begin{figure}
\begin{center}\vskip-1.5cm
\includegraphics[angle=0,width=\columnwidth]{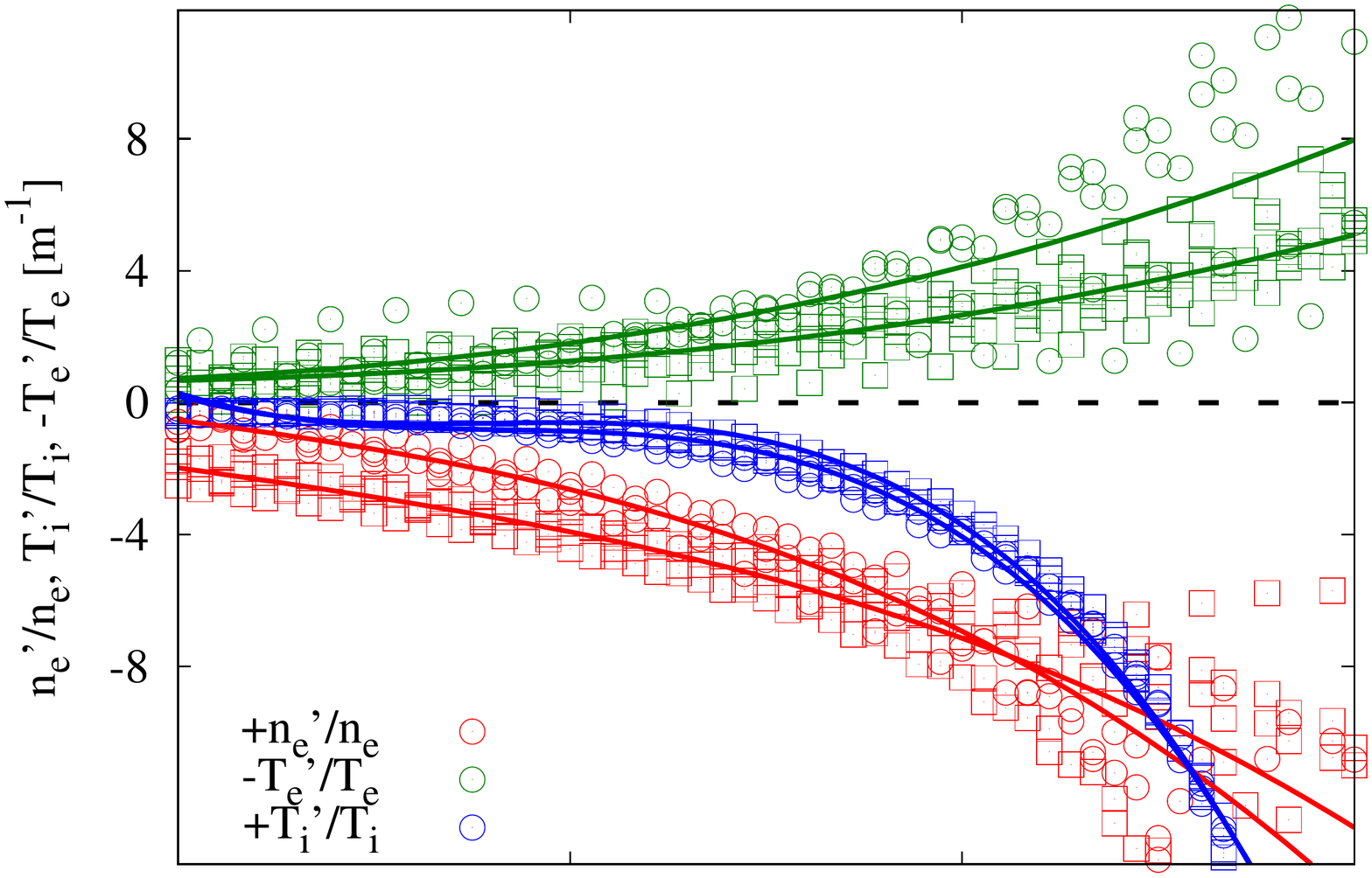}\vskip-3.5cm
\includegraphics[angle=0,width=\columnwidth]{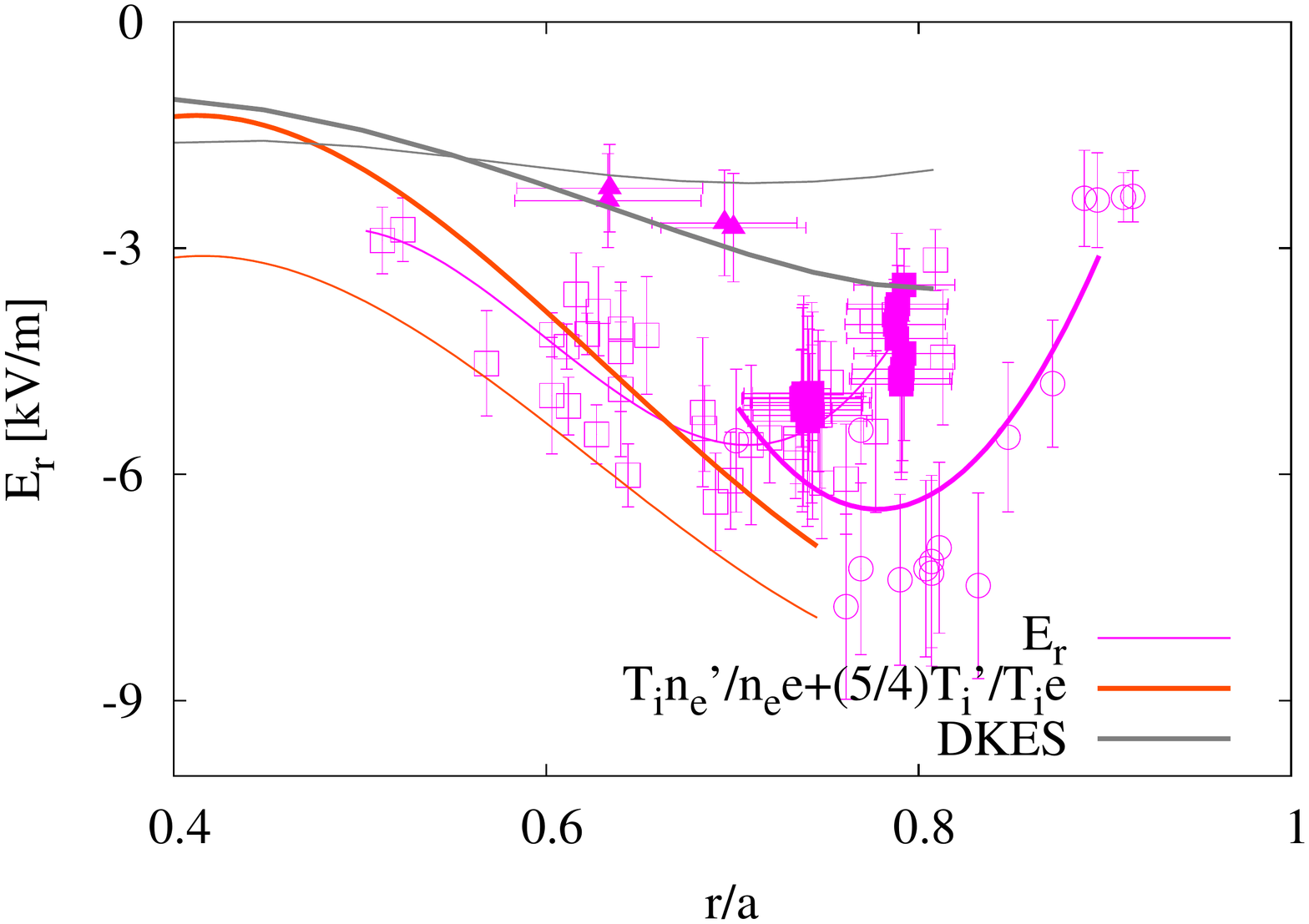}\vskip-3.5cm
\end{center}
\caption{Top: thermodynamical forces associated to the density, electron temperature and ion temperature gradients of TJ-II discharges for high density (open circles) and intermediate density (open squares). Bottom: measured radial electric field of TJ-II discharges for high density (open circles), intermediate density (open squares), new discharges with intermediate density (closed squares) and low density (closed triangles). {For reference we plot fits to the old measurements (magenta), the ion root estimates (orange), and simulations with DKES (grey); thicker line correspond to higher density}.}\label{FIG_RESULTS_TJ-II}
\end{figure}

\begin{figure}
\begin{center}\vskip-2.5cm
\includegraphics[angle=0,width=0.6\columnwidth]{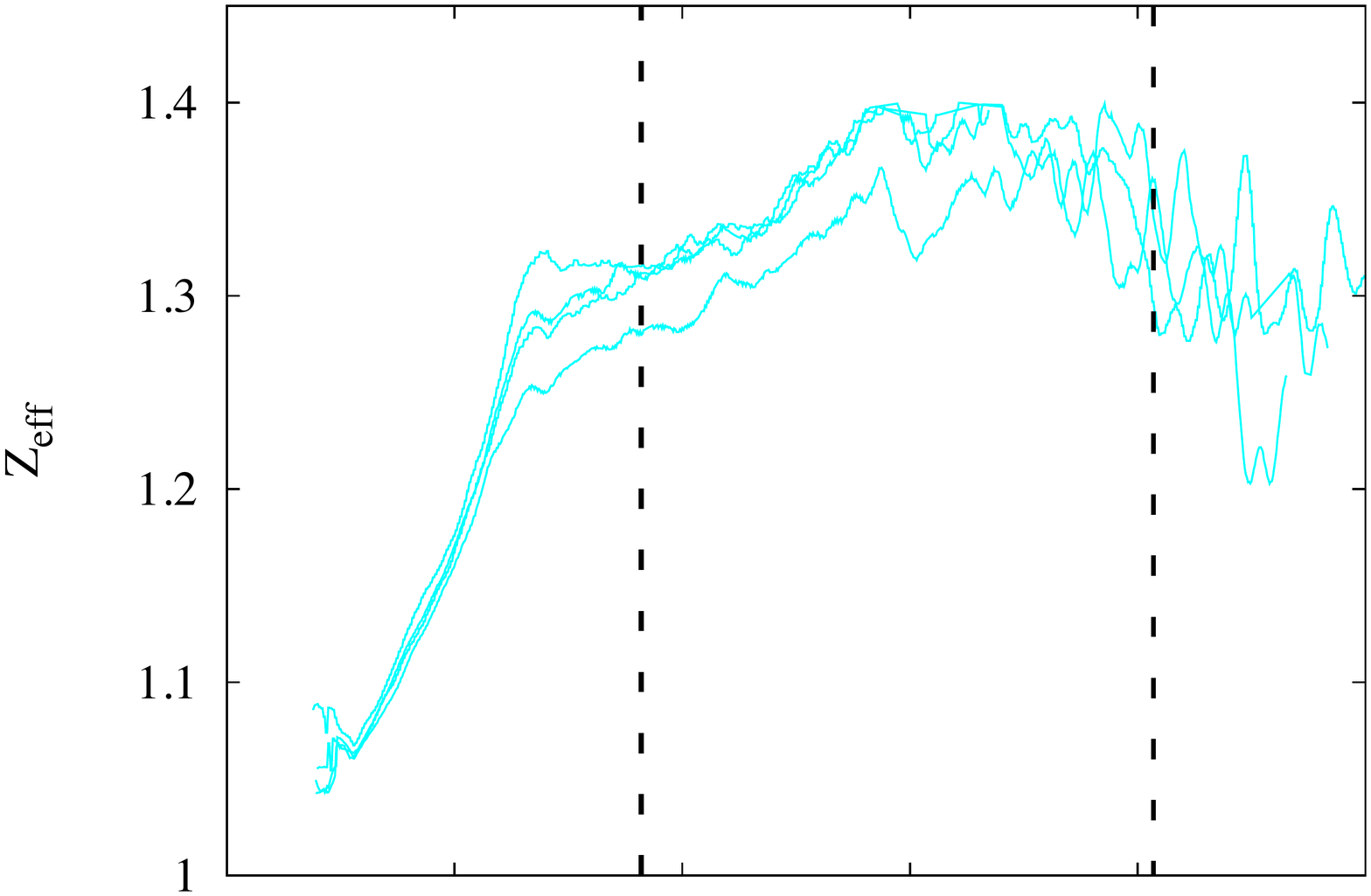}\vskip-2.2cm
\includegraphics[angle=0,width=0.6\columnwidth]{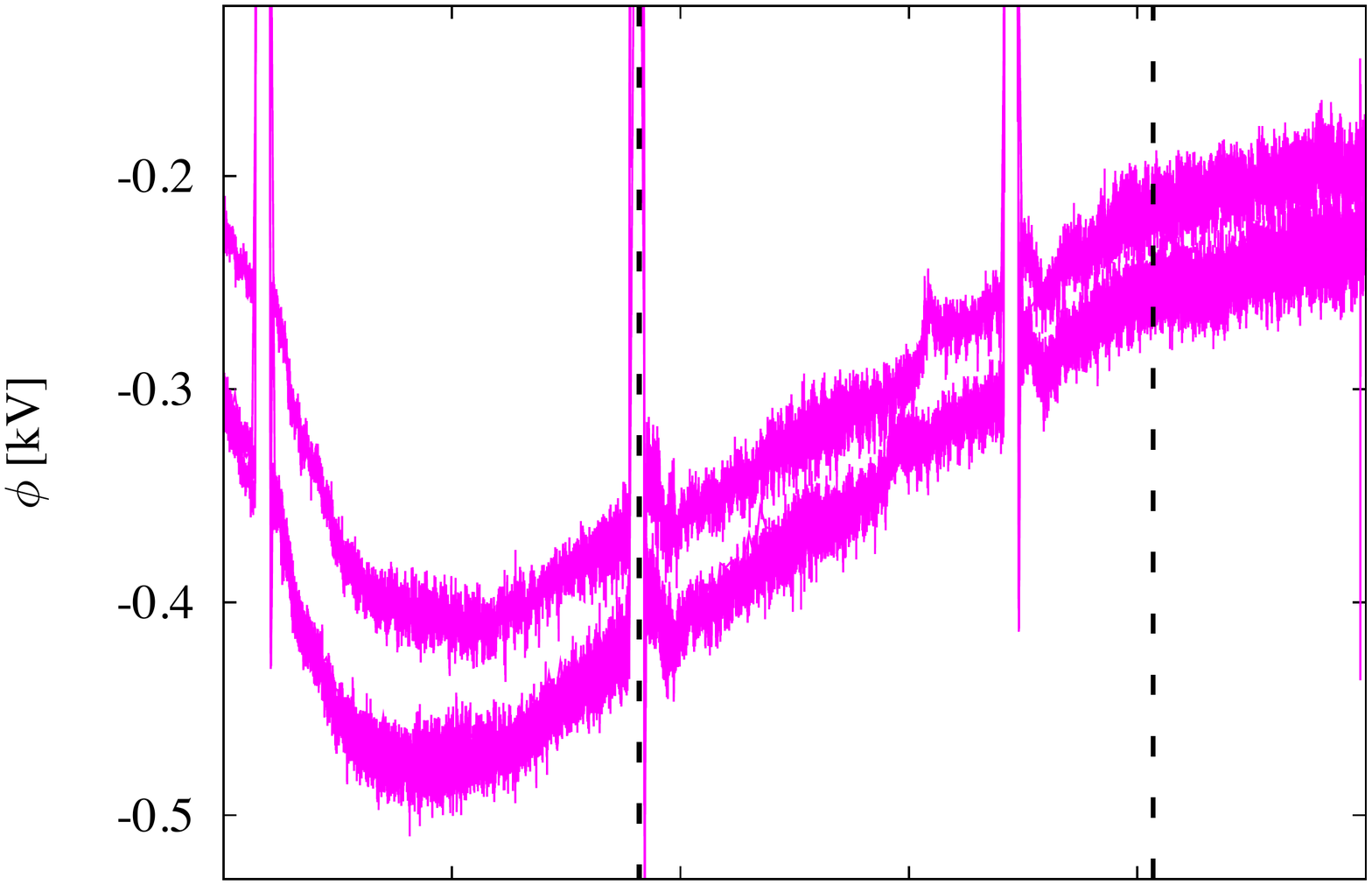}\vskip-2.2cm
\includegraphics[angle=0,width=0.6\columnwidth]{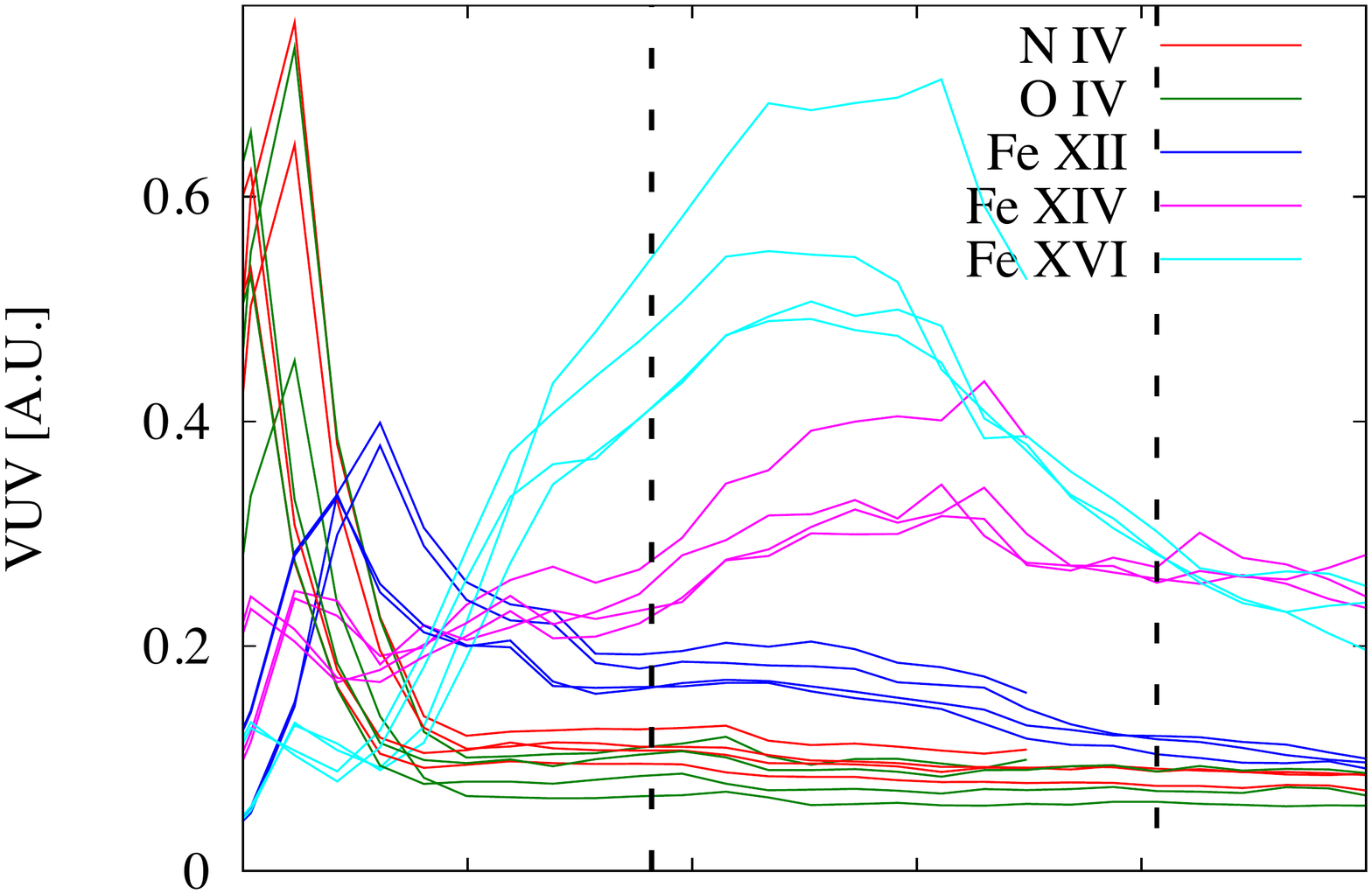}\vskip-2.2cm
\includegraphics[angle=0,width=0.6\columnwidth]{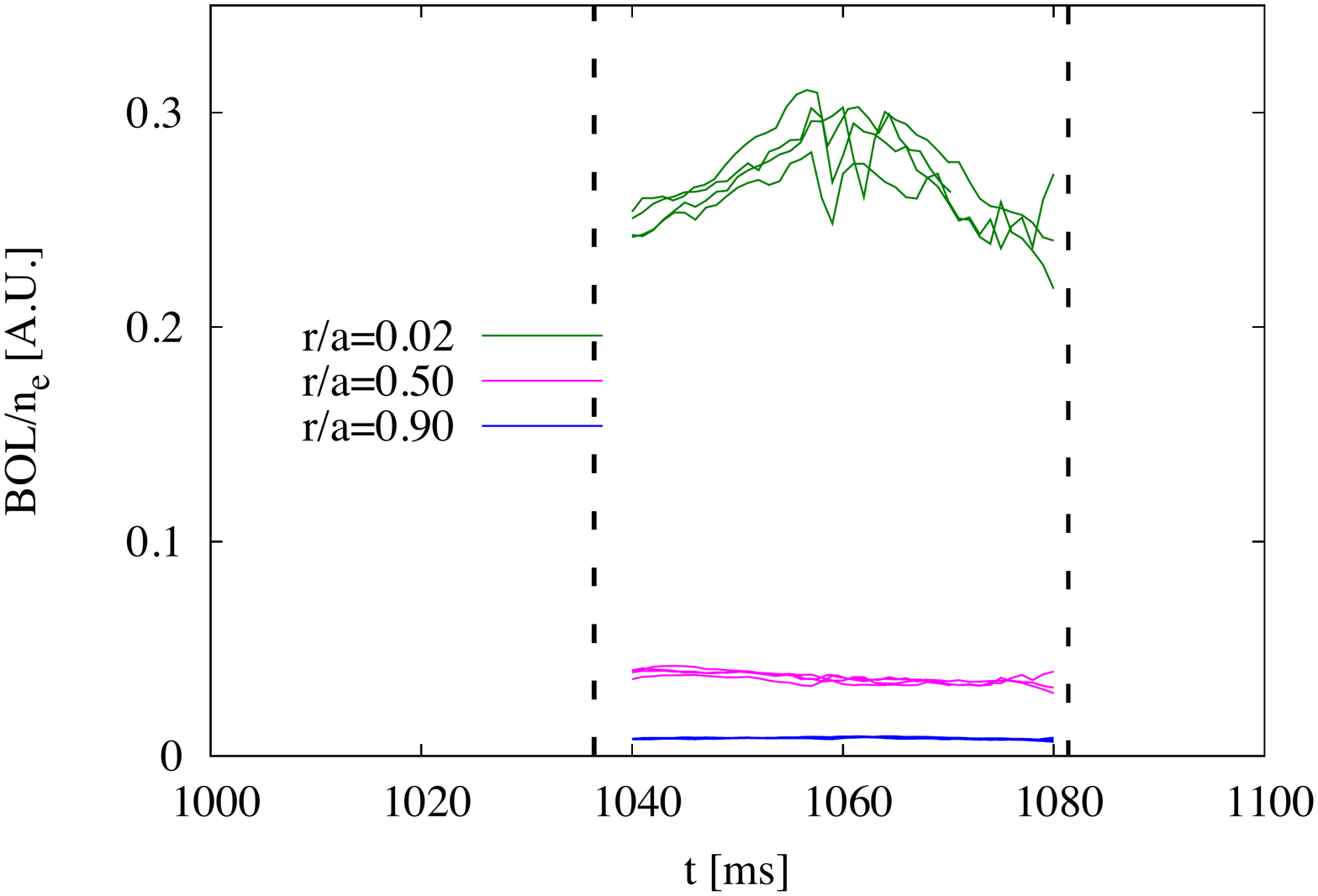}\vskip-4.5cm
\end{center}
\caption{{Time traces for TJ-II discharges \#41706, \#41709-41711, from top to bottom: effective charge, core plasma potential, total emission from several impurity ions, and central and edge radiation (from tomographic reconstruction of bolometry signal) normalized by the average density. Vertical lines indicate the TS time.}}\label{FIG_TRACES_TJ-IIb}
\end{figure}

 \section{TJ-II}\label{SEC_TJ-II}

In order to learn from inter-machine comparison, we now study the case of low density plasmas of TJ-II heated with Neutral Beam Injection (NBI). We first note in figure~\ref{FIG_DIA} (bottom) that, although lower than for LHD, the ion temperatures necessary to have small radial electric field are very large; for the plasma parameters accessible to TJ-II, the thermal ions are in the plateau regime. Nevertheless, the contribution to transport of ions of higher energy in the Maxwellian distribution, which are in the $1/\nu$ regime (for $E_r\!=\!0$) or $\sqrt{\nu}$ regime, is not negligible, as can be seen from the non-linearity of $\Gamma_i(E_r)$ e.g. in~\cite{velasco2013ncldt}. Furthermore, the electron collisionality is lower than that of the ions.

Twelve new discharges (\#41700-41711) are created as shown in figure~\ref{FIG_TRACES_TJ-II} for four of them (\#41706, \#41709-41711): plasmas are heated by NBI only (counter-injection for $t\!<\!1020\,$ms, co-injection for $t\!>\!1020\,$ms) and the line-averaged density (measured with a microwave interferometer) is reduced after the initial increase. This is achieved thanks to good wall-conditioning~\cite{sanchez2009transitions} and low gas puffing. The core electron and ion temperatures (measured by Electron Cyclotron Emission (ECE)~\cite{delaluna2001ECE} and Neutral Particle Analyzer~\cite{fontdecaba2014NPA} respectively) are kept fairly constant {around $T_i(r/a<0.3)\approx 0.15\,$keV}. Since ion temperature gradients are not available for these new discharges {(only core temperature values, without spatial resolution, are)}, we also recall \textit{old} plasmas from~\cite{arevalo2014flows2} {with similar core ion temperature} (18 discharges from data-sets \#28257-28263, \#29199-29208, \#31096-31110, \#32246-32247, \#32304-32306; from these, \#28263 is part of the validation activity~\cite{satake2015ishw} within the ISHPDB), where $T_i$ up to $r/a\!=\!0.7$ was measured using CXRS. Since these plasmas are close to the cut-off density, the ECE measurements are not reliable at the beginning of the discharge, and we will focus in the time interval ${1040}\,$ms $\!<\!t\!<\!1080\,$ms; we will nonetheless show the time traces for $t\!<\!{1040}\,$ms, as they will prove useful when discussing the behaviour of impurities.

Figure~\ref{FIG_PLASMAS_TJ-II} shows fits~\cite{milligen2011bayes} to the density and electron temperature profiles measured with Thomson Scattering (TS)~\cite{herranz2003TS} and Helium Beam diagnostic~\cite{hidalgo2004hebeam} and the ion temperature profile of the selected discharges. In the new discharges, the electron profiles are measured at the times marked with vertical lines in figure~\ref{FIG_TRACES_TJ-II}, and only core ion temperature is available. We see that the old discharges can be grouped into two sets: high and intermediate density (in the gradient region); the new discharges can be grouped into two sets: intermediate (around $r/a\!=\!0.6$) and low density. In all cases, differences in the temperatures between datasets are much smaller. This is more clearly seen in figure~\ref{FIG_RESULTS_TJ-II} (top), which shows the relevant thermodynamical forces of these discharges.

In figure~\ref{FIG_RESULTS_TJ-II} (bottom) we compare the measured radial electric field with the ion root estimate (up to $r/a\!=\!0.7$, since the ion temperature profile at outer radial positions is an extrapolation) {and DKES simulations}. We note that good agreement was found in the core between predictions of radial electric field from DKES and CXRS measurements of poloidal rotation~\cite{arevalo2014flows2} (and no difference between co- and counter-injection); nevertheless, large error bars in the measurements prevent us from studying its parameter dependence using that data, so we focus on the gradient region, where measurements from Doppler reflectometry~\cite{happel2009doppler} were also available, {although the quantitative comparison with DKES is worse}. Good agreement is observed in the measured radial electric field between old and new discharges corresponding to equivalent densities. We see that, for highest densities, $E_r \approx T_in_e'/n_ee + 5T_i'/4e$, showing again that the ambipolar equation could be described with ions in the $\sqrt{\nu}$ regime and neglecting the electron contribution. For intermediate densities, the radial electric field becomes less negative, {both in the experiment and in the DKES simulations}, even though the thermodynamical forces do not change (the one related to the density gradient even increases slightly), an indication of the same effect that we have discussed for LHD. For the new discharges of lowest density, the radial electric field is even closer to zero.

Finally, it may be interesting now to see whether this change in radial electric field, although small, has an effect on impurity transport. We show in figure~\ref{FIG_TRACES_TJ-IIb} an estimate of the volume-averaged effective charge $Z_{eff}$ from X-rays~\cite{medina1999xrays}. Since $Z_{eff}\equiv \sum_{b\neq e}n_bZ_b^2/n_e$, $Z_{eff}-1$ can be considered proportional to $n_I/n_e$. A clear change is observed in its time evolution around $t\!=\!1060\,$ms, for similar values of the density for the four discharges: for averaged densities higher than $0.8\!\times\!10^{19}\,$m$^{-3}$, it grows; for lower densities, it slightly decreases. This behaviour could be due to two causes: either a change in transport (that could be connected to the reduction of the absolute value of $E_r$) or to a reduction in the impurity source: generally speaking, lower densities mean lower particle flux to the wall and smaller impurity source.

In order to look for a connection with a change in transport, we measure with a vacuum ultraviolet spectrometer (VUV)~\cite{mccarthy2010VUV}  the time evolution of the total content for three specific impurities: oxygen and nitrogen (charge-state $Z_I\!=\!3$ in both cases), concentrated mainly in the edge, and iron (charge-states $Z_I\!=\!11$, $Z_I\!=\!13$, and $Z_I\!=\!15$), mostly present in the core region. Without spatial resolution, we cannot account for the ionization and recombination terms in the evolution of these impurities, and it is difficult to estimate changes in transport: for instance, the evolution of the different lines in the early part ($t\!<\!1020\,$ms) of the discharges is obviously dominated by ionization and recombination, as the electron temperature rises during the plasma start-up. Nevertheless, in between (${1040}\,$ms $\!<\!t\!<\!1080\,$ms), the electron temperature is roughly constant {in the core}, and we will find some observations that are clearly consistent with the prediction of reduced impurity influx.

The first point to be considered is that the NBI system can be an additional source of iron through two contributions: beam interaction of the counter-NBI with the duct~\cite{liniers2013duct} and shine-through neutrals from both NBIs colliding with the vacuum vessel~\cite{guasp1999shinethrough}; the former should constitute a constant source of iron up to $t\!=\!1020\,$ms, while the latter should cause an increasing iron source during the length of the discharge, since lower density leads to increased shine-through. From the fact that the iron content keeps increasing tens of milliseconds after the counter-NBI has been switched-off {(the confinement time for iron in low-density ECH-heated plasmas of TJ-II is between 5 and 10$\,ms$~\cite{zurro2014blow})}, while the density is decreasing, we infer that the shine-through has a measurable contribution to the iron source; the observed change in the VUV signal around $t\!=\!1060\,$ms would then be caused by transport. 

Secondly, a connection can be made with the variation of the radial electric field: impurities of higher charge state are the ones that follow more closely the time evolution of $Z_{eff}$. This would indicate that it is transport what explains the time evolution of the impurity content. In order to support this interpretation, we now return to equation~(\ref{EQ_LINEAR}) and we set $\Gamma_I\!=\!0$, thus obtaining a impurity density distribution given by:
\begin{equation}
n_I(r)=n_I(a)\exp{\int_a^{r_0}\mathrm{d}r \frac{Z_IeE_r(r)-T_i'(r)}{T_i(r)}}\,.\label{EQ_SSI}
\end{equation}
This is the steady-state distribution towards which the impurity density is evolving (in the absence of sinks and sources{, and anomalous contribution to the impurity flux}). If we compare two situations $t_1$ and $t_2$ characterized by the same temperature $T_i$, the same impurity edge density $n_I(a)$, and different electron density (and hence radial electric field), the relation between the steady-state impurity densities at a given radial position $r_0$ should be:
\begin{equation}
\frac{n_I(r_0,t_2)}{n_I(r_0,t_1)} =\exp{\int_a^{r_0}\mathrm{d}r \frac{Z_Ie[E_r(r,t_2)-E_r(r,t_1)]}{T_i(r)}} \approx \exp{\left(\frac{Z_Ie[\Phi(r_0,t_1)-\Phi(r_0,t_2)]}{T_i(r_0)}\right)}\,,\label{EQ_RAT}
\end{equation}
where $\Phi$ is the electrostatic potential. {Here we have made a number of assumptions: the second equality is only valid for flat ion temperature profile; a steady state is not reached within the duration of a typical NBI TJ-II discharge, and  $n_I(a)$ is not measured; finally, a turbulent contribution to the impurity flux in equations~(\ref{EQ_SSI}) and (\ref{EQ_RAT}) can not be ruled out, specially close to the edge}. Nevertheless, equation~(\ref{EQ_RAT}) provides some general indication: a less negative potential $\Phi$ at $r_0$ (caused by a less negative value of $E_r$ at $r\!>\!r_0$) should be connected to a reduction of the impurity density at $r_0$. In particular, if $E_r$ becomes less negative at all radial positions, the change in impurity density should be stronger in the core. We have seen this kind of behaviour with the VUV. In order to support this connection, we show in figure~\ref{FIG_TRACES_TJ-IIb} the evolution of the core plasma potential measured with a Heavy Ion Beam Probe (HIBP)~\cite{bodarenko2001HIBP} and the core radiation obtained from tomographic inversion of the signal of bolometer arrays~\cite{ochando2006bolo}. This quantity depends on the electron temperature and (approximately linearly) on the electron density. Therefore, for constant electron temperature, if normalized by the line-averaged density, it can be associated to the local impurity density. The plasma potential is negative {and its absolute value decreases monotonically in time, reaching values closer to zero than those usually measured in TJ-II NBI plasmas (see e.g. figure 5 of~\cite{melnikov2007hibp})}; the core radiation, similarly to what happens to the effective charge, shows a change in behaviour below certain density. {We note that no change is observed in the radiation from the edge, from which we infer that the impurity edge density $n_I(a)$ has not changed significantly, as we assumed in equation~(\ref{EQ_RAT}) (since the global parameters of the discharge had been modified, the impurity content at the edge might have been different and this effect might have had an impact on the core impurity density, recall equation~(\ref{EQ_SSI})).} Let us finally note that a spike is observed at $t\!=\!1060\,$ms, at all radial positions, and it is also observed in the electron temperature. It could be connected to small changes in the magnetic configuration: the toroidal current $I_t\!\sim\!kA$ changes during the duration of the discharge, so does the rotational transform, and low-order rationals may enter or exit the plasma. Rationals are known to modify the local electric field~\cite{bodarenko2010rat,estrada2016rat}, but the event detected by bolometry is global. Rationals aside, the effect of a change of rotational transform in the radial electric field is expected to be negligible in ion root~\cite{velasco2012er}, as compared with the changes measured in figure~\ref{FIG_RESULTS_TJ-II}.

\section{Discussion}\label{SEC_DIS}

In this work we have discussed plasmas of very low ion and electron collisionalities. We have shown with experiments and simulations in several devices that these plasmas tend to present a small and negative radial electric field (which can even become positive and large if the collisionality decreases further~\cite{yokoyama2002lhd}), and that this behaviour of the bulk plasma may have a measurable impact on impurity transport. It should be emphasized that these plasmas are different from those usually termed as CERC~\cite{yokoyama2007cerc}: in those plasmas, $T_e\!\gg\!T_i$ (hence $\nu_e^*\!\ll\!\nu_i^*$) was the main cause for the $1/\nu$ electron flux become larger than the $\sqrt{\nu}$ flux of the ions, overcoming the difference in Larmor radius, and thus leading to the onset of a very positive radial electric field. In these plasmas, the differences in transport regime, and also the reduced ion transport due to the favourable scaling of the $\sqrt{\nu}$ with ion temperature play a key role.

More experimental and theoretical efforts are required for a better description of impurity transport in helical devices. But the qualitative picture allows to postulate that high temperature of bulk species may be a key ingredient to avoid impurity accumulation in plasmas such as those with impurity hole, even if the radial electric field is negative, contrary to usual expectations. It also {shows a mechanism through which intense core heating can contribute to impurity control and ash removal}.

Finally, since bulk transport is well described by theory-based models, these results can be used for {discussing the feasibility of} optimization strategies in helical reactors. Indeed this intermachine study provides additional support to previous validations (e.g.~\cite{nagaoka2015etb}) of theoretical predictions~\cite{yokoyama2002lhd} and gives confidence on its extrapolation to larger devices. {In order to show that high ion temperature in the core of helical devices is not fundamentally incompatible with low core impurity content, we can first} compare the derived parameter dependence:
\begin{eqnarray}
T_i \sim \epsilon^0 R^{4/5}B^{2/5}n_e^{2/5}\,, \label{EQ_FSCA2}
\end{eqnarray}
where we have rearranged equation~(\ref{EQ_FSCA}) and kept only the configuration parameter-dependence, with the ISS04 empirical stellarator scaling for energy confinement time~\cite{yamada2005taue}: 
\begin{eqnarray}
T_i \sim \frac{\tau_E P}{n_e} \sim \epsilon^{2.28}R^{2.92}B^{0.84}n_e^{-0.46}P^{0.39}\,,
\end{eqnarray}
where $P$ is the input power (we note that the ISS04 scaling has been found in agreement with scenario predictions based on neoclassical transport with electrons and ions in the $1/\nu$ and $\sqrt{\nu}$ regimes respectively~\cite{dinklage2007taue}). The zero-dimensional parameters of the magnetic configuration have the same sign in both scalings, which indicates that changing them in order to improve energy confinement time (e.g., going towards larger devices) also increases the temperature at which good impurity behaviour is expected. However, the exponents of the energy confinement scaling are larger. This would suggest that both good energy confinement and good impurity and helium ash behaviour could possibly be obtained at the same time.

{Nevertheless, even if there is no fundamental incompatibility, the temperature required for achieving this regime may be practically too high. In order to make an assessment, we start from the experimental parameters of the impurity hole regime of LHD ($R\!=\!3.5\,$m, $B\!=\!3\,$T, $n_e\!=\!2\times 10^{19}\,$m$^{-3}$, $T_i\!=\!5\,$keV). Helical reactor designs based on the experimental devices LHD, W7-X, NCSX (National Compact Stellarator Experiment) have minor radii between 8 and 20$\,$m, and toroidal magnetic field on axis around 5$\,$T~\cite{sagara2010reactor}. Using equation~(\ref{EQ_FSCA2}) we see that the regime could be achieved for $n_e\!=\!2\times 10^{19}\,$m$^{-3}$ and $T_i$ between 12 and 25$\,$keV \changes{(neoclassical optimization could slightly increase these temperatures, as it happens within the LHD configuration space)}. While this range of temperatures is close to being optimal for maximizing fusion power, as estimated using the Lawson criterion~\cite{lawson1955criteria}, the density of typical reactor scenarios is one order of magnitude higher~\cite{sagara2010reactor}. Therefore, in order to combine the benefits from high density operation with those of impurity hole plasmas, a complete understanding of the latter is yet required. Specifically, depending on how they scale with the plasma parameters, mechanisms such as asymmetries in the impurity distribution on the flux-surface or turbulent processes could help to reduce this density gap.}

\section*{Acknowledgments}

The authors are indebted to C Hidalgo and A Melnikov for their help with the HIBP data, and to B. Liu for useful discussions about previous experiments at TJ-II. This work has been carried out within the framework of the EUROfusion Consortium and has received funding from the Euratom research and training programme 2014-2018 under grant agreement No 633053. The views and opinions expressed herein do not necessarily reflect those of the European Commission. The International Stellarator-Heliotron Database is pursued under the auspices of IEA Implementing Agreement for Cooperation in Development of the Stellarator-­Heliotron Concept (2.10.1992) and the authors are indebted to all its contributors and previous responsible officers, and in particular to A Dinklage. This research was supported in part by grants ENE2012-30832 and ENE2013-48679, Ministerio de Econom\'ia y Competitividad, Spain, by the International Hubs for Natural Science Research program, National Institutes of Natural Sciences, Japan, and the NIFS Collaboration Research program NIFS15KNST079.

\end{document}